\documentclass{article}
\usepackage[utf8]{inputenc}
\usepackage{graphicx}
\usepackage[top=1in,left=1in,right=1in,bottom=1in]{geometry}
\usepackage{amsmath}

\usepackage{xcolor}
\usepackage{dcolumn}
\usepackage{amsmath}
\usepackage{amsfonts}
\usepackage{bm}
\usepackage{epstopdf}
\usepackage{picture}
\usepackage{enumitem}
\usepackage{afterpage}
\usepackage{placeins}
\usepackage{float}
\usepackage{enumerate}
\usepackage[utopia]{mathdesign}

\usepackage{xr}
\usepackage{caption}
\usepackage{mathtools}
\begin{document}

\title{Epsilon-Near-Zero (ENZ)-based Optomechanics}
\author{Y. Kiasat$^{1,}$\footnote{These authors contributed equally to this work.} , M. G. Donato$^{2,*}$, M. Hinczewski$^{3,*}$, M. ElKabbash$^{3}$, T. Letsou$^{3}$,\\ R. Saija$^{4,}$\footnote{Email: rsaija@unime.it, onofrio.marago@cnr.it, gxs284@case.edu, engheta@ee.upenn.edu} ,  O. M. Marag\`{o}$^{2,\dag}$, G. Strangi$^{3,5,\dag}$, and N. Engheta$^{1,\dag}$}
\date{}
\maketitle
\noindent
$^1$Department of Electrical and Systems Engineering, University of Pennsylvania, Philadelphia, PA 19104, USA.\\
$^2$CNR-IPCF, Istituto per i Processi Chimico-Fisici, I-98158 Messina, Italy.\\
$^3$Department of Physics, Case Western Reserve University, 10600 Euclid Avenue, 44106, Cleveland, Ohio, USA. \\
$^4$Dipartimento di Scienze Matematiche e Informatiche,
Scienze Fisiche e Scienze della Terra, Universit\`a di Messina, I-98166 Messina, Italy.\\
$^5$CNR-NANOTEC Istituto di Nanotecnologia and Department of Physics, University of Calabria, Via Pietro Bucci 87036, Rende, Italy\\
\\

\textbf{Optomechanics deals with the control and applications of mechanical effects of light that stems from the redistribution of photon momenta in light scattering. Here, we investigate, analytically and numerically, optical forces on polarizable particles in proximity of epsilon-near-zero (ENZ) metamaterials. We look at the general features of the repulsive-attractive optomechanics from the nano to the microscale exploiting different theoretical methods (dipole approximation, finite elements calculations, transition (T-)matrix). We discuss the role of realistic layered materials, as our ENZ substrate, on optical forces and analyze the influence of composition and shape by studying a range of complex particles (dielectric, core-shell, plasmonic ellipsoids). Physical insights into the results are discussed and future research directions are forecasted. Our results provide new possibilities in exploiting engineered materials and surfaces for the manipulation and tailoring of light-induced forces in optomechanics.}
\newpage

\section*{Introduction}
Recent developments in the field of metamaterials and metasurfaces have provided useful platforms for manipulating and tailoring light-matter interaction with numerous applications ranging from cloaking \cite{alu2005achieving, schurig2006metamaterial}, enhanced spontaneous emission \cite{alu2009boosting}, sensing \cite{alu2009cloaking}, signal processing and information handling \cite{silva2014performing, estakhri2019inverse}, and nonreciprocity \cite{coulais2017static}, just to name a few. Among various classes of metamaterials, the epsilon-near-zero (ENZ) and near-zero-index (NZI) structures have attracted increasing attention due to their unique features in light-matter interaction \cite{silveirinha2006tunneling, liberal2017near, reshef2019nonlinear, kinsey2019near}. In such structures, relative permittivity and/or relative permeability attain values near zero, thus making the effective refractive index of the structure near zero.  Consequently at the operating frequency, the wavelength in these media is “stretched”, making the phase of the signal approximately uniform across this structure \cite{engheta2013pursuing}.  As a result, the waves exhibit “static-like” spatial distributions, while temporally they are dynamic. This has led to numerous exciting wave phenomena, with several potential applications \cite{silveirinha2006tunneling, liberal2017near, reshef2019nonlinear, kinsey2019near, liberal2017rise}. One such feature is the possibility of levitation of electrically-polarized nanoparticles in the vicinity of ENZ substrates \cite{Rodriguez2014}. In our earlier work, we theoretically showed that an infinitesimally small nanoparticle, when electrically polarized at a given frequency, could be levitated when placed near an ENZ substrate. This phenomenon, which was inspired as a classical analogue to the Meissner effect (levitated magnets in proximity of superconductors), can provide a new approach in optomechanics when manipulation of electrically polarizable particles is desired in the presence of optical fields.

Careful manipulation of particles with light, which has a long history dating back to the pioneering work of Ashkin in 1970s \cite{Ashkin1970,AshkinAPL71}, has played important roles in various areas, from biology \cite{Fazal2011} to nanoscience and nanotechnology \cite{Marago2013}. At the nanoscale, various methodologies have been used for such optical manipulation, including trapping \cite{Spesyvtseva2016,JuanNP11}, pushing \cite{GargiuloNL16,Donato2018}, and binding \cite{DemergisNL12,Donato_NL_2019}, with different materials such as dielectrics, semiconductors, plasmonic, and biological \cite{Jones2015}. The surrounding media can be vacuum, air, or liquid. Optical tweezing \cite{Ashkin1986,Jones2015} is usually achieved using optical beam shaping to generate desired potential traps \cite{Dholakia2011}. Recently, new approaches to optical manipulation of objects without a beam-shaping were proposed. Soljacic and co-workers \cite{ilic2017topologically} proposed that the motion of a Janus particle with spatially asymmetric absorption can be controlled by changing the incident wavelength. Ilic and Atwater \cite{ilic2019self} proposed self-stabilizing optical manipulation of macroscopic objects by controlling the anisotropy of the scattered light from the structured object's surface. Both approaches, however, rely on structuring the object in lieu of the incident light.

In the present work, we merge the two fields of ENZ metamaterials and of optical trapping, providing a new platform, which we name ENZ-based optomechanics, for manipulating and controlling mechanical motion of particles in vicinity of ENZ structures. We explore, numerically and analytically, how various parameters, such as the size, shape and composition of the particle and its distance to the ENZ substrate affect the optomechanical forces on such particles. We consider both homogeneous and layered structures as our ENZ substrates.  In recent years, researchers have been able to tailor the effective permeability and permittivity of composite media by engineering the electric and magnetic resonances of nanostructures. Together with related developments in nanophotonics, metamaterials provide unprecedented freedom to define and sculpt electromagnetic modes. Metamaterials allow to alter the topology of photonic isofrequency surfaces - which govern the momentum and energy of optical modes inside a medium - contrarily to conventional bounded spherical and ellipsoidal isofrequency surfaces in natural dielectrics\cite{krishnamoorthy2012topological}. Among many other extreme optical features, unbounded iso-frequency surfaces in hyperbolic dispersion metamaterials \cite{poddubny2013hyperbolic} and point-like vanishing surfaces in epsilon-near-zero (ENZ) media  \cite{mahmoud2014wave} constitutes two examples of advanced modal engineering.
In particular, epsilon-near-zero metamaterials provide extended modes with uniform phase over micrometer length scales inducing profound effects on nanoscopic light-matter interactions. These deeply subwavelength structured surfaces support unique electromagnetic modes that can be used in sub-diffraction imaging, \cite{jacob2006optical} and waveguiding \cite{silveirinha2006tunneling}, spontaneous emission engineering \cite{cortes2012quantum} and biosensing \cite{sreekanth2016extreme}.

In the following, we introduce the geometry of the problem, and discuss the electromagnetic modeling for the structure, along with the dipole approximation. We also present extensive numerical results, based on the finite-element method (using the commercial software COMSOL Multiphysics\textsuperscript{\tiny\textregistered}) and on the T-matrix methods. We also present a series of results for various parameters involved in the problem. Physical insights into the results are presented and future directions are discussed.

\begin{figure}
\includegraphics[width=\textwidth]{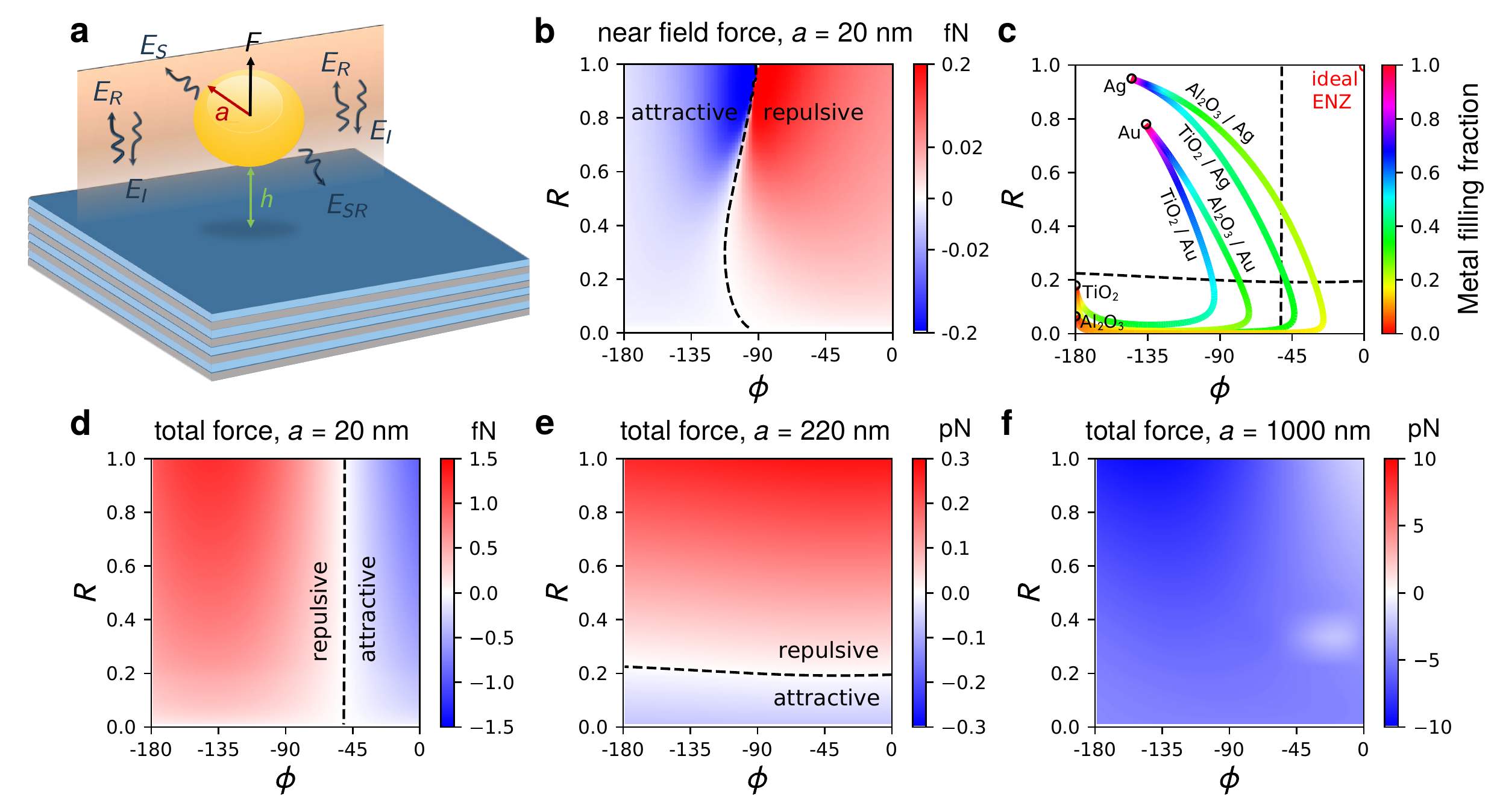}
\caption{Geometry and optical force parameter space maps.  a) Sketch of the geometry. We consider a generic particle, in principle of any shape and composition, in front of a metamaterial surface at an edge-to-edge distance, $h$, immersed in an external medium (e.g., water) of refractive index $n_{\rm m}$. The origin of the coordinate system is placed on the surface so that the $z$ axis is positive in the semi-infinite space where the particle resides. A monochromatic optical wave illuminates the particle and surface at normal incidence so that the total field is the superposition of incident ($E_{\text{I}}$), reflected ($E_{\text{R}}$) and scattered fields ($E_{\text{S}}$, $E_{\text{SR}}$). The resulting optical force can be either attractive (negative) or repulsive (positive). b) Near-field force $(R,\phi)$ map. We explore the range of reflectivity, $R$,  and phase, $\phi$, related to the reflection of the incident wave on a generic surface. The ideal ENZ surface ($R = 1$, $\phi = 0$)) is in the top right corner of the map and shows a repulsive force. Here the near-field force component is calculated in the dipole approximation for a polystyrene (dielectric constant $\varepsilon_{p}$=2.543 at 560 nm)) particle of radius $a=20$ nm at a fixed edge-to-edge distance of $h=10$ nm in water. c) Curves of $(R,\phi)$ for different substrates consisting of alternating metal and dielectric layers, in the thin layer limit where effective medium theory is valid.  The color of each curve indicates the metal filling fraction.  The zero force lines from panels d and e are superimposed as dashed lines.  d-f) Total force $(R,\phi)$ maps as calculated from T-matrix methods for a dielectric particle size of $a=20$ nm (d), $a=220$ nm (e), and $a=1000$ nm (f), respectively, and at a fixed $h=10$ nm. The total force maps have a structure that is strongly dependent on particle size. This is due to the increase of the scattering force component that, for large particles, overcomes any other gradient-like force component that is dominant for nanoparticles.}
\label{Fig1}
\end{figure}

\paragraph{Geometry of the problem.}
Figure \ref{Fig1}a presents the geometry of our problem. A polarizable particle, made of a single nonmagnetic material (or multilayered materials), surrounded by an external medium (e.g., water) of refractive index $n_{\rm m}$, is located at an edge-to-edge distance $h$ above a metamaterial substrate. The particle can be spherical (or other shapes as will be discussed later in the manuscript), and it is made of a (dielectric or metallic) material with a relative permittivity $\varepsilon_{\rm p}$. The substrate can be considered as a homogenized nonmagnetic medium with relative permittivity near zero at the frequency of operation, or a layered structure engineered to function as ENZ.

A monochromatic optical wave is illuminating this structure at normal incidence. The goal is to evaluate the optical force on the particle and to investigate how various parameters, radius $a$, edge-to-edge distance $h$, particle’s permittivity, and signal frequency affect the optical force's magnitude and direction, \textit{i.e.} whether it is a repulsive (positive) or an attractive (negative) force. In the next section, in order to gain some physical insight we start by assuming the polarizable particle to be represented by an infinitesimally small electric dipole, and discuss the analytical approach for evaluating the force acting on this particle.  In the subsequent sections, we will expand our approach to include the full-wave numerical simulations of the problem, allowing to consider realistic sizes and shapes for this particle.

%\section*{Methods} - Briefly describe theory methods and background
\paragraph{Dipole approximation.}
We first consider a particle size much smaller than the light wavelength ($a << \lambda$) so that optical forces can be calculated analytically within the dipole approximation (DA) \cite{Chaumet2000, AriasGonzalezJOSAA03, Jones2015}. Due to its simplicity, the dipole approximation can provide useful results that can be compared with more complex light scattering approaches (T-matrix, finite elements methods) at the nanoscale \cite{Polimeno2018}.
\begin{figure}
\begin{center}
\includegraphics[scale=0.5]{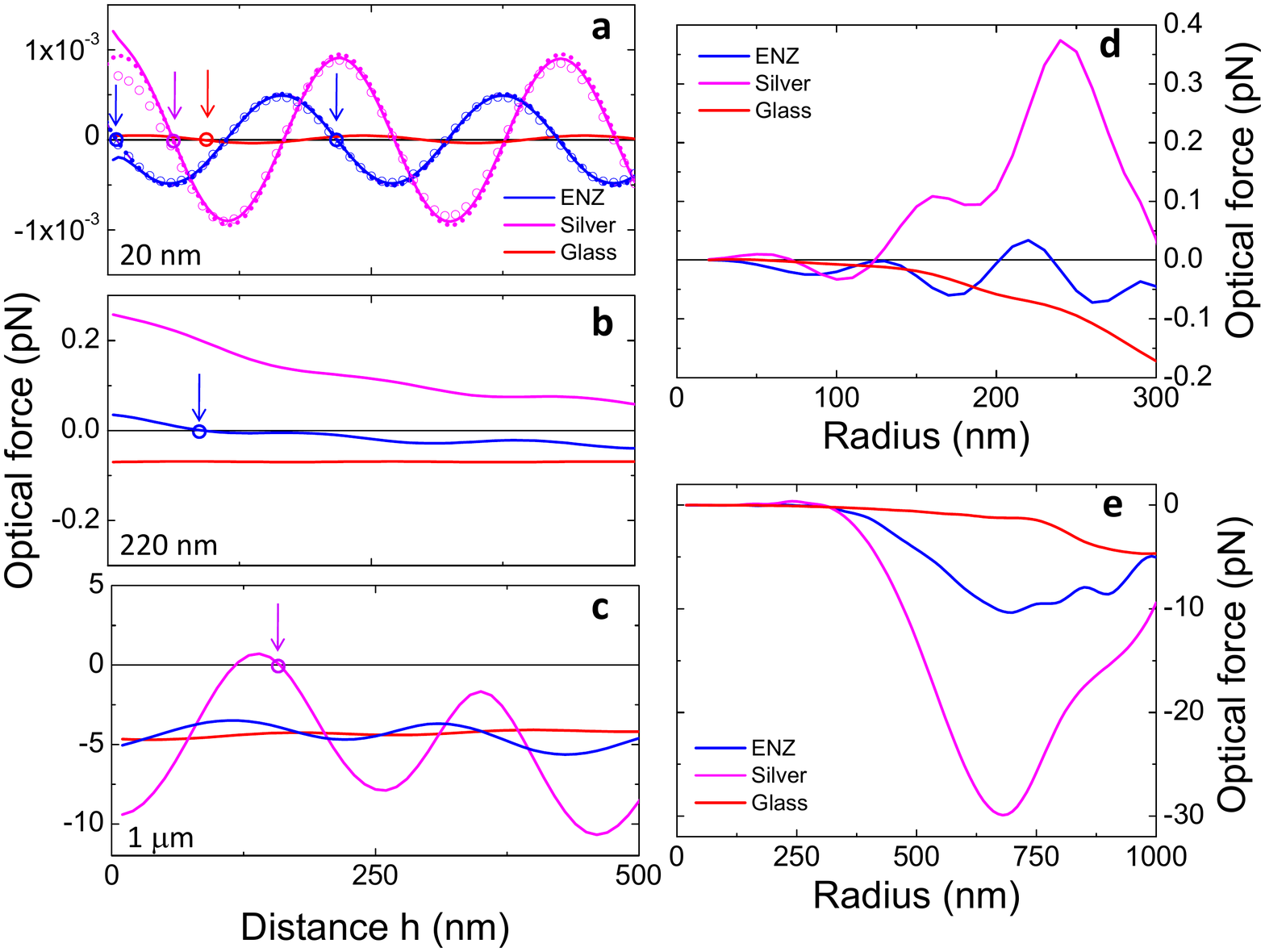}
\caption{Total optical force as a function of edge-to-edge distance, $h$, for polystyrene particles of different size: a) $a=20$ nm), b) $a=220$ nm, and c) $a=1$ $\mu $m. Different approaches for the calculation of the force are compared: dipole approximation (short dots), COMSOL (circles) and T-matrix (continuous lines). Different surfaces , glass (red), silver (magenta), ENZ (blue), yield very different optomechanical interactions in terms of force amplitude, modulation with respect to $h$, and phase-shifts. Arrows indicate self-binding points, where particles are stably trapped in front of the surface. (d,e) Size dependence of the total optical force for fixed edge-to-edge distance, $h$=10 nm. For small particles the gradient force component of the partially reflected plane wave dominates, resulting in a dependence with particle size, while for large particles radiation pressure has a major contribution resulting in a negative force pushing the particle towards the surface.}
\label{Fig2}
\end{center}
\end{figure}

We start our analysis from the near-field force component. It has been shown \cite{Rodriguez2014} that in front of an ENZ surface an emitting point dipole source is subjected to a near-field repulsive force, reminiscent of the Meissner effect in superconductors \cite{Rodriguez2014}. This portion of the force, which we refer to as the ``near-field'' force, is due to the interaction of the emitting dipole with the substrate  (excluding the force due to the presence of the incident and reflected waves).  When all forces are considered (including the forces caused by the incident and reflected waves), the forces are called ``total force''. We can extend the result of Ref. \cite{Rodriguez2014} to a finite-sized polarizable particle illuminated by an incident field by considering the radiated power upon scattering, $P_{\rm rad}=\sigma_{\rm scat} I(z)$, in terms of the scattering cross section, $\sigma_{\rm scat}$, and light intensity, $I(z)$. Thus, the near-field force component is (see Supp. Info.):

\begin{equation}\label{Fenz_pw1}
\mathrm{F_{enz}}(z)\approx - \frac{9}{512 \pi^4 c} \mathrm{Re}\left( \frac{\varepsilon_{\rm s}-\varepsilon_{\rm m}}{\varepsilon_{\rm s}+\varepsilon_{\rm m}} \right) \left( \frac{\lambda}{n_{\rm m} z} \right)^4 \sigma_{\rm scat} I(z)
\end{equation}

\noindent where $z$ is the axial coordinate ($z=h+a$, $a$ is the radius of the particle, $h$ is the edge-to-edge distance of the particle from the surface), $c$ is the vacuum speed of light, $\varepsilon_{\rm m}=\varepsilon_0 n_{\rm m}^2$ is the permittivity of the surrounding medium, $n_m$ is the refractive index of the medium, and $\varepsilon_{\rm s}$ is the complex dielectric permittivity of the ENZ surface.

In Fig. \ref{Fig_bead_all}, a panel summarizing the results of the calculation of the %total optical force (a-c) and of the
near-field force (d-f) on a 20 nm dielectric bead in water is shown. Three different surfaces are considered: lossless (Im$(\varepsilon_s)$=0), with medium loss (Im$(\varepsilon_s)$=0.5), and with high loss (Im$(\varepsilon_s)$=0.8). The comparison with the results obtained for a point dipole in vacuum \cite{Rodriguez2014} shows that in this work the presence of a medium (water) broadens the repulsive near field force region from $-1<\varepsilon_s<1$ to $-1.77<\varepsilon_s<1.77$; moreover, as already observed\cite{Rodriguez2014}, even in surfaces with high loss there is still a repulsive near-field force.

In order to explore how the ENZ surface can influence the near-field and total forces on the particle, we evaluate such effects in terms of the amplitude $\rho$ and phase $\phi$ of the complex reflection coefficient of an incident wave from this surface. In Fig. \ref{Fig1}b, the near-field force on a $a$=20 nm radius dielectric bead at $h$=10 nm from the surface has been calculated as a function of the surface reflectivity $R$=$\left|\rho e^{i\phi}\right|^2=\rho^2$ and phase angle $\phi$ which are connected to the surface complex refractive index  $\tilde{n}= n_{\rm s}+i k_{\rm s}$ by \cite{Born1980_6th_ed}:

\begin{align}
\rho=\sqrt{\frac{(n_{\rm m} - n_{\rm s})^2 + k_{\rm s}^2}{(n_{\rm m} + n_{\rm s})^2 + k_{\rm s}^2}} &&  \phi=\arctan \left[ \frac{-2n_{\rm m} k_{\rm s}}{n_{\rm m}^2 - n_{\rm s}^2 - k_{\rm s}^2 }\right]
\label{R and phi}
\end{align}

Here we use $e^{- i\omega t}$ as our time harmonic convention. The calculated near-field force can reach a fraction of femtonewton for an incident intensity of approximately 5.6$\cdot 10^8$ W/m$^2$ (corresponding to a typical experimental configuration, see Sect. 1.2 of the Suppl. Info.) and changes character from attractive to repulsive when the reflection phase angle changes from $\phi$=-$\pi$ to $\phi=0$. Metals such as Au or Ag, having a certain amount of absorption ($k_s$ in Eq. \ref{R and phi}), are in the attractive region of the near-field force (compare Figs. \ref{Fig1}b and c). On the contrary, in front of an ideal ENZ surface, having $R$=1  and $\phi$=0, the near-field force is repulsive.   Substrates of alternating metal and dielectric layers can span a broader range of $\phi$ and $R$ values.  In the limit of layers much smaller than the incident wavelength, where effective medium theory (EMT) is valid, we show the $(R,\phi)$ results for four different metal / dielectric mixtures in Fig. \ref{Fig1}c (see Sect. 4 of the Supp. Info. for more details on the EMT calculation).  The metal filling fraction is indicated by the color of the curve.  Depending on the fraction, we can switch the sign of the force from attractive to repulsive and vice versa.  If we go beyond EMT and take into account the finite thickness of layers in real structures, as described in the discussion of Fig.~\ref{Fig3} below, we can achieve an even wider range of $\phi$ and $R$ values.

We now consider the total optical force from an incident field on a nanoparticle calculated in DA. This is the sum of  two main components: a gradient force, $\mathrm{F}_{\rm grad}$, and a scattering force, $\mathrm{F}_{\rm scat}$ \cite{Jones2015}. For plane wave illumination (for Gaussian beams see Sect. 1.2 of Supp. Info.) impinging normally to the ENZ surface, the force components are influenced by incident and reflected fields. Thus, considering only the axial direction $z$, they are written as (see Supp. Info.):

\begin{equation}\label{Fgrad_pw1}
\mathrm{F_{grad}}=\frac{1}{2}\frac{n_{\rm m}}{c \varepsilon_{\rm m}} \mathrm{Re}(\alpha) \frac{d I(z)}{dz}
\end{equation}

\begin{equation}\label{Fscatt_pw1}
\mathrm{F_{scat}}=\frac{n_{\rm m}}{c}\sigma_{\rm ext}I_0\left[ \rho^2 - 1\right]
\end{equation}

\noindent where $\alpha$ is the particle complex polarizability \cite{DraineAJ93},

\begin{equation}\label{alpharad}
\alpha=\frac{\alpha_{0}}{1-i\frac{k^3\alpha_0}{6 \pi \varepsilon_{\rm m}}}
\end{equation}

\noindent $\alpha_0$ is the Clausius-Mossotti polarizability and $\sigma_{\rm ext}=\frac{k}{\varepsilon_{\rm m}}\mathrm{Im}({\alpha})$ is the extinction cross-section, related to the particle absorption and scattering \cite{AriasGonzalezJOSAA03,Jones2015}, with $k=2\pi n_{\rm m}/\lambda$ the wave number and $\lambda$ the wavelength.

The gradient force, $\mathrm{{F}}_{\rm grad}$, drives the particle towards the maximum (minimum) of the modulated light intensity profile for positive (negative) real part of the polarizability. On the other hand, the scattering force, $\mathrm{{F}}_{\rm scat}$, is constant with respect $z$, and it always pushes the particle along the beam propagation direction.\\

In Fig. \ref{Fig1}d, the $(R,\phi)$ map of calculated total axial force on a polystyrene $a$=20 nm bead (dielectric constant $\varepsilon_{p}$=2.543 at 560 nm) at $h$=10 nm from the surface is shown. The total force is one order of magnitude larger than the near-field force (Fig. \ref{Fig1}b) and shows a change in the repulsive-attractive character when the phase angle changes from $-\pi$ to 0 , respectively. This is due to the gradient force (see also Fig. \ref{Fig2}a, short dots) that dominates the optomechanical response and drives the particle towards the high field intensity regions. The change of the phase of the reflection coefficient shifts the intensity modulation resulting from the interference between the incident and reflected wave. Thus, the high intensity points shift accordingly and the sign of the force changes around $\phi\sim -\pi/4$.

We now calculate the total optical force on the dielectric bead in front of glass, Ag and ENZ surfaces as a function of distance, $h$. The ENZ material is chosen so that $n_{s}\approx 0.476$ and $k_{s}\approx 0.511$, in order to obtain a  real part of complex permittivity close to zero and an imaginary part close to 0.5 to include unavoidable losses of realistic systems. This choice leads to values of $R$ and $\phi$ similar to those in the experimentally fabricated layered substrates described below, corresponding to the point marked with a red star in Fig.~\ref{Fig3}. The strong modulation resulting from the standing wave is clearly visible. The points with zero force and negative slope are trapping points that correspond to equilibrium positions for the particle dynamics (arrows in Fig. \ref{Fig2}a). For the case of the ENZ surface the equilibrium point closest to the surface occurs at $h\sim$10 nm, while for the glass and Ag surface they occur at $h\sim$ 85 nm and $h\sim$ 60 nm, respectively. By linearizing the force at the equilibrium points, $F(z) \approx -\kappa z$, a trap spring constant $\kappa$ can be calculated. The trap spring constants $\kappa_{\mathrm{ENZ}}$ and $\kappa_{\mathrm{Ag}}$ calculated in front of ENZ and Ag surfaces can be compared to the spring constant $\kappa_S$ calculated, in a standard single-beam optical tweezers setup, with the same particle and at the same light intensity (see Supp. Info.). The spring constants are $\kappa_{\mathrm{ENZ}}=15$ fN/$\mu$m and $\kappa_{\mathrm{Ag}}=27$ fN/$\mu$m, while the trap spring constant $\kappa_S$ in a standard optical tweezers setup is two order of magnitude lower, $\kappa_S=0.24$ fN/$\mu$m. The beneficial effect of the ENZ and Ag reflective surfaces on the trapping is evident.
The increasing size of the particles corresponds to larger optical forces and different trapping points (see Fig. \ref{Fig_dielbeads_vsradius} in Supp. Info. for DA calculations on larger size nanoparticles at 50 and 100 nm).

\paragraph{Full-wave simulations.}
In order to calculate optical forces on larger particles, we use two different full-wave modeling approaches based on the transition (T-)matrix formalism \cite{waterman1971symmetry,Borghese2007book} and on finite-elements methods using the commercial software COMSOL Multiphysics\textsuperscript{\tiny\textregistered}, respectively.
In particular, electromagnetic scattering from particles near to or deposited on a plane surface that separates two homogeneous media of different optical properties in the T-matrix formalism \cite{Borghese2007book,JOSA95,JOSA99,AO99} can give account on the role of the different multipoles in the particle-surface interaction (see Supp. Info. for details). Indeed, the presence of the surface can have a striking effect on the scattering pattern from the particles, because the field that illuminates the particle is partly or totally reflected by the surface and the reflected fields contribute both to the exciting and to the observed field.
Moreover, the field scattered by the particle is reflected by the interface and thus contributes to the exciting field. In other words, there are multiple scattering processes between the particles and the interface. As a result, the field in the accessible half-space includes the incident field $\mathbf{E}_{\rm I}$, the reflected field from the interface $\mathbf{E}_{\rm R}$ (as we would have if no particle were present), the scattered field from the particle $\mathbf{E}_{\rm S}$ and, finally, the field that after scattering by the particle is reflected by the surface, $\mathbf{E}_{\rm SR}$, related to $\mathbf{E}_{\rm S}$
by the reflection condition (see Fig. \ref{Fig1}a). Thus, the observed field, superposition of  $\mathbf{E}_{\rm S}$ and $\mathbf{E}_{\rm SR}$, includes all the scattered and scattered-reflected multipole contributions (see Supp. Info. for more details).

It is possible to define the T-matrix for particles in the presence of the interface that is the starting point to calculate optical forces and torques either by direct integration of the Maxwell stress tensor (MST) over a closed surface containing the particle \cite{Jones2015} or by exploiting the general expressions of optical force and torque in terms of multiple expansion \cite{Saija2005,Borghese2006,Borghese2007}. The optical force is obtained in COMSOL by direct integration of the MST that is calculated based on the total electric field and magnetic field which include the incident fields, the scattered fields by the particle, and all the reflected fields by the surface (see Supplementary Information for details on full wave methods).

The results obtained in DA from the different surfaces are compared in Fig. \ref{Fig2}a with those obtained by using full electromagnetic calculations based on COMSOL (circles), and T-matrix methods (continuous lines). A very good agreement is clearly observed. In all approaches, the total optical force on small particles is modulated by the sinusoidal term of the gradient force. Its magnitude is larger (in the fN range) on more reflective surfaces and its phase changes sign going from a Ag to an ENZ substrate, leading to the formation of optical trapping points at different distances (arrows in Fig. \ref{Fig2}a). In brief, the gradient force dominates the ENZ-optomechanics for small particles even in proximity of the surface.

T-matrix and COMSOL allow the calculation of optical forces for larger particles than in DA. In
Fig. \ref{Fig1}e and f, the $(R,\phi)$ map of total axial force calculated with the T-matrix
approach on a $a$=220 nm bead (Fig. \ref{Fig1}e) and $a$=1 $\mu$m bead (Fig. \ref{Fig1}f), at
$h$=10 nm from the surface, are shown. The comparison with Fig. \ref{Fig1}d highlights the strong
dependence of the total optical force on the bead size. The repulsive-attractive behaviour is
driven by the competition between gradient force and scattering force, which may give repulsive
behaviour, for intermediate size beads, in surfaces having large reflectivity (see Fig.
\ref{Fig1}e); however, at large bead size (Fig. \ref{Fig1}f), scattering force overcomes gradient
force, and the total optical force is attractive in front of every type of surface.

\begin{figure}
\begin{center}
\includegraphics[scale=1.15]{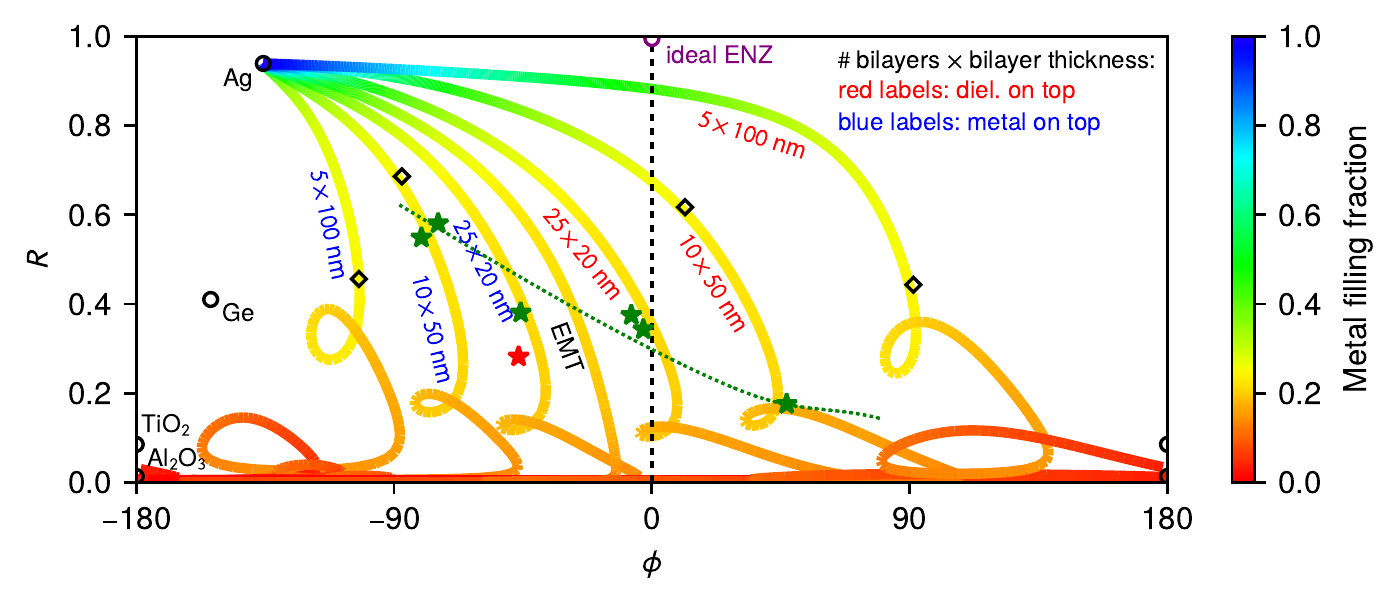}
\caption{Accessing the full range of reflectance ($R$) and reflected phase ($\phi$) via layered metamaterials.  $R$ versus $\phi$ is illustrated for light at normal incidence with wavelength $\lambda = 560$ nm reflected from the surface of a thin film stack. The thick curves of varying color are transfer matrix numerical calculations labeled $n \times d$, where $n$ refers to the number of bilayers in the stack, and $d$ the thickness of each bilayer.  The total thickness $n d = 500$ nm is kept constant.  The bilayers consist of individual Ag and Al$_2$O$_3$ layers, with the fraction of Ag in the bilayer indicated by the metal filling fraction color.  The red $n \times d$ labels correspond to systems where the dielectric is the upper layer in each bilayer (the one closest to the surface), while the blue labels are the ones where the metal is on top. Curve labeled EMT is the effective medium approximation to the system, which corresponds to $n \to \infty$, $d \to 0$ with $n d = 500$ nm.  In all the above cases the superstrate is water and the substrate is glass.  For comparison we show points indicating the $R$ and $\phi$ values for a simple interface between a water superstrate and a pure material substrate (Ag, Au, Ge, TiO$_2$, Al$_2$O$_3$, and an ideal ENZ).  We also show experimental results (green stars, details in the Supp. Info.) involving a water superstrate and 5 trilayers (Al$_2$O$_3$/Ag/Ge from top to bottom, where Ge is present as a thin wetting layer to ensure fabrication quality). The dotted green trend line corresponds to keeping the Ag and Ge layer thicknesses fixed at 15 nm and 2.5 nm respectively, while varying Al$_2$O$_3$ thickness from 80 nm to 20 nm (left to right). In order to compare EMT calculation with the full-wave analysis, COMSOL is used (black diamonds) to calculate $R$ for different layered structures with various metal's filling fraction (0.4 and 0.6), layer's thickness (50 nm and 100 nm) while keeping the total thickness of the layered structure unchanged (500 nm) and also different order of material in the stack (metal on top, blue labels, and dielectric on top, red labels).}
\label{Fig3}
\end{center}
\end{figure}

In Fig. \ref{Fig2}b and c the T-matrix calculations of the total optical force on larger particles are shown as a function of the edge-to-edge distance from ENZ, Ag and glass surfaces. The larger size of these particles with respect to the nanosized bead in Fig. \ref{Fig2}a highlights the increased contribution of the scattering force on the gradient force. The scattering force is detrimental towards stable equilibrium positions in front of glass surface for the 220 nm radius bead and in front of both ENZ and glass surfaces for 1 $\mu$m radius bead. The lower reflectivity of these surfaces as compared to the reflection from the Ag surface does not allow an efficient balance between scattering force from incoming and reflected beams, increasing the scattering force contribution with respect gradient force and hindering the trapping.

 In Fig. \ref{Fig2}d and \ref{Fig2}e the results are reported for increasing bead size at fixed distance, $h=$10 nm, from the ENZ, Ag or glass surfaces. It is shown that at small bead size (below approximately 300 nm radius), the gradient force modulates the total force. At increasing bead size, the particle extinction cross section increases, consequently the scattering force is predominant on the gradient force, inhibiting equilibrium points and inducing an effective attractive force directed towards the surfaces.

\paragraph{Epsilon-Near-Zero Metamaterials.}
Regarding layered ENZ materials, we have demonstrated experimentally that it is possible to control the optical topology and to induce the ENZ behavior by designing and fabricating subwavelength layered lattice structures as a result of interlocking noble metals and dielectric thin films \cite{sreekanth2013experimental}. Upon selecting metal-dielectric bilayers, the thickness of each layer, the filling fraction and the number of bilayers,  the frequency of the optical topological transition in the iso-frequency surface leading to the epsilon-near-zero behavior can be tailored.
The lattice structure is fabricated as a five tri-layer system using $\mathrm{Al_2O_3}$, Ag, and Ge from top to bottom. The Ag layer thicknesses were in the range of 10-25 nm, with a thin Ge layer (1-3 nm) underneath to ensure surface wetting. The $\mathrm{Al_2O_3}$ layer thicknesses were systematically varied between roughly 20 nm and 80 nm across different material systems (Fig. \ref{Fig3}), subsequently tuning the frequency of the topological transition.
In previous studies we used effective medium theory to calculate the dielectric permittivity of the entire structure, as opposed to more recent inverse design approaches to account for a wider material parameters space. We perform spectroscopic ellipsometry measurements to evaluate the dielectric tensor components and the dispersive behavior of the layered structure. By fitting the measured angular reflectance and the ellipsometry parameters $\psi$ and $\Delta$, we can directly obtain the effective optical constants of the multilayer slab. Using the transfer matrix method, we can then predict the magnitude and phase of reflection at normal incidence with a water superstrate.  The green stars in Fig.~\ref{Fig3} represent these predicted values from 6 samples consisting of a 5 bilayer Al$_2$O$_3$/Ag thin-film stack with a Ge seed layer to ensure the uniformity of the Ag films. By varying the thickness of the $\mathrm{Al_2O_3}$ layers, we covered a phase range of $\Delta\Phi$  $\approx$ 180$^\circ$  and  reflectance range of $\Delta R$ $\approx$ 0.5.  The full range of accessible $R$ and $\phi$ values is even larger if we expand the design space of the substrate to include different numbers of bilayers and metal filling fractions.  The thick curves in Fig.~\ref{Fig3} show transfer matrix calculations of $(R,\phi)$ for Al$_2$O$_3$/Ag stacks with different structural parameters indicated by the labels.  In all cases the total thickness of the stack was kept fixed at 500 nm.  The color at each point along the curves corresponds to the metal filling fraction.  In the limit of many thin bilayers we approach the EMT result of Fig.~\ref{Fig1}c, which is also reproduced here for comparison. Note that actual layered materials can achieve positive values of $\phi$, while homogeneous materials (for example those described by EMT) are confined to the $\phi < 0$ subspace.
%\textbf{To do:  update the green experimental points to reflect water and not air superstrate.}

\begin{figure}
\begin{center}
\includegraphics[width=\textwidth]{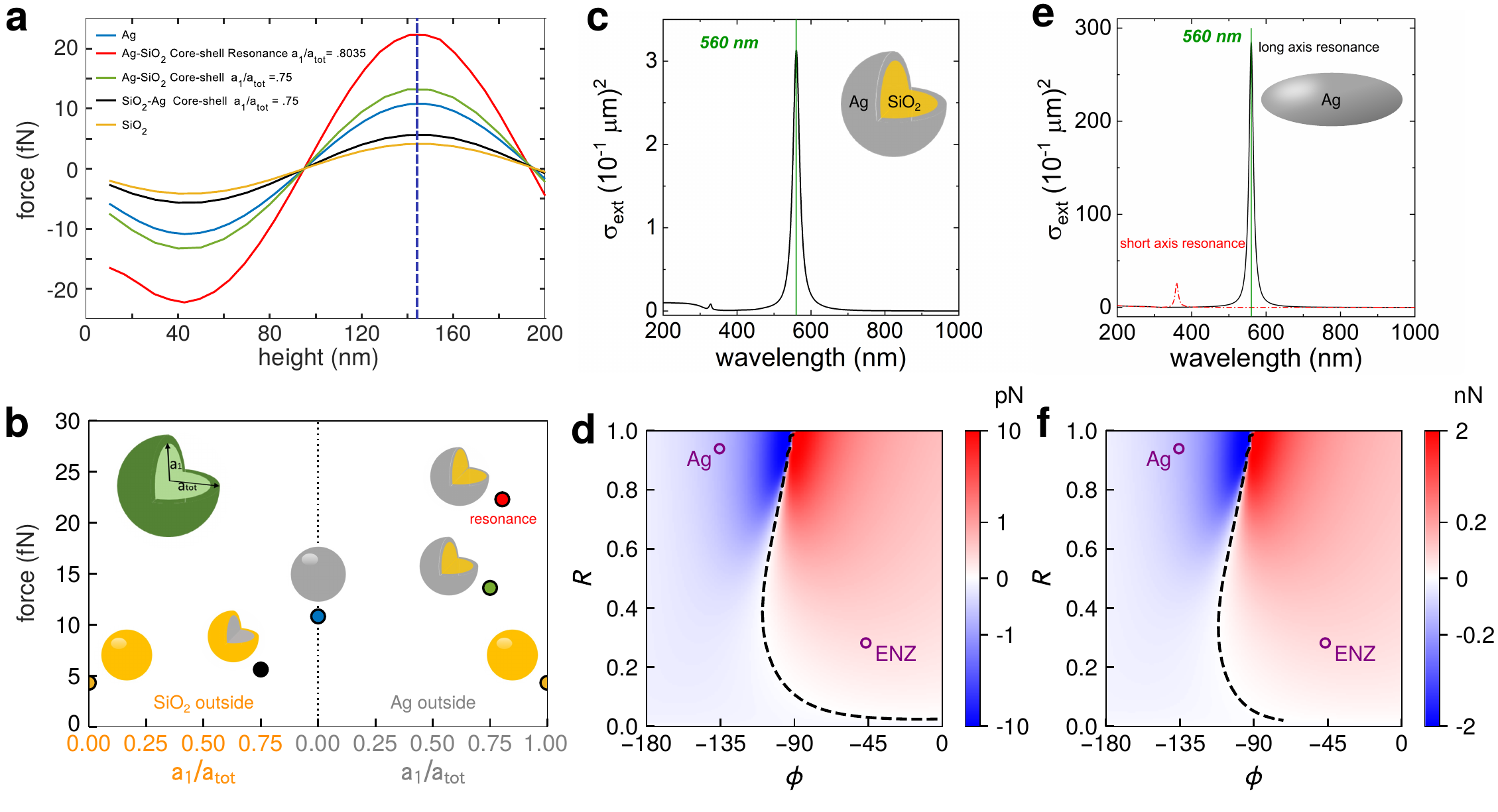}
\caption{Role of polarizability and shape on optical forces. a) COMSOL calculation of total optical force on core-shell particles based on SiO$_2$ and Ag as a function of the distance from the metamaterial surface. Different ratios between core radius $a_1$ and total particle radius $a_{\rm tot}$=20 nm have been considered. Moreover, both materials have been considered as a core. b) Maximum force found in each core-shell structure considered, as pointed out by the dashed blue line in a).
c) Extinction spectrum of the SiO$_2$-Ag core-shell particle (total radius $a_{\rm tot}$=20 nm and core radius $a_{1}$ =16.1 nm) in water.
d) $(R,\phi)$ contour plot of the total optical force for the core-shell particle at $h$=10 nm from the surface. The ENZ and Ag surfaces used for the calculation of the optical forces in DA approximation are shown as circles.
e) Extinction spectra of Ag prolate ellipsoid in water oriented with the long axis parallel to the field (black solid line) and oriented with the short axis parallel to the field (red dashed line). The resonances relative to the long and short axes are indicated.
f) $(R,\phi)$  contour plot of the total optical force on the Ag ellipsoid at $h$=10 nm distance from the surface. In the calculation, the spheroid is aligned with the long axis in the direction of the wave polarization. The ENZ and Ag surfaces used for the calculation of the optical forces in DA approximation are shown as circles. The optical force is in the order of tens of pN in front of ENZ (repulsive) and Ag (attractive). The force can be close to 200 pN if the spheroid is in front of an ideal ENZ surface, having R=1 and $\phi$ =0.}
\label{Fig4}
\end{center}
\end{figure}

\paragraph{Complex particles (core-shell, ellipsoids, ENZ)}
In addition to spherical beads, we evaluate the optical forces on different types of particles in front of  dielectric, metallic or ENZ surfaces. We consider spherical core-shell particles based on $\mathrm{SiO_2}$ and Ag, an Ag prolate spheroid and a spherical particle made by an ENZ material.

We first used COMSOL simulation to calculate the forces on core-shell structures in front of layered ENZ material at 560 nm illumination. The total particle radius $a_{\rm tot}$ is fixed at 20 nm. The particles had alternatively $\mathrm{SiO_2}$ or Ag as the core, with the other material as the shell. In Fig. \ref{Fig4}a the total force on the core-shell particles as a function of the distance from the ENZ surface is shown. It is clearly observed that the presence of Ag in the outer shell enhances the total force with respect of the inverse structure having $\mathrm{SiO_2}$ as the shell, but also with respect to the pure Ag sphere. The highest value of the force is found (red curve in Fig. \ref{Fig4}a) for a $\mathrm{SiO_2}$-Ag core-shell structure having a core radius of $a_1=16.1$ nm and an Ag shell 3.9 nm thick which, as shown in Fig. \ref{Fig4}c, is at the resonance condition at the ENZ wavelength.

%To enhance the optical force, we used a $\mathrm{SiO_2}$-Ag core-shell particle designed to be resonant at approximately 560 nm and having a total radius $a_{tot}$= 20 nm. The calculation of the extinction cross-section shows (Fig. \ref{Fig4}a) that the resonance condition is fulfilled  if the core-shell structure has a core radius $a_1=16.1$ nm, and the Ag shell thickness is 3.9 nm.

As shown in Figures \ref{Fig4}d, \ref{Fig_cs}d and \ref{Fig_cs}e, the particle resonance at 560 nm enhances the optical force to the piconewton range (Fig. \ref{Fig4}d) but only at very short distances from the surfaces, being repulsive in the ENZ case (Fig. S\ref{Fig4}d) and attractive in the Ag case (Fig. \ref{Fig_cs}e). Otherwise, the total optical force is at the fN range.

Specifically, at the resonance $\mathrm{F_{enz}}$ is in the piconewton range close to the ENZ
surface (from $h$=0 nm to roughly 10 nm). The gradient force, $\mathrm{F_{grad}}$, has an
oscillating character, but its amplitude is smaller ($\approx$ 1 fN) than $\mathrm{F_{enz}}$, due
to the small real part of the polarizability at resonance ($\mathrm{Re(\alpha)=0.04 \cdot 10^{-32}
\ F m^2}$). On the contrary, $\mathrm{F_{scatt}}$ is large (tens of femtonewton), because of the
large extinction cross section at resonance. Thus, at 560 nm (black curve in Fig.
\ref{Fig_cs_vs_wl}a), the total force is repulsive and in the piconewton range close to the
surface, but becomes attractive and approximately constant as the $\mathrm{F_{enz}}$ contribution
fades off with distance.

The behaviour of the forces on the core-shell particle can also be studied for wavelengths smaller
and larger than the particle plasmon resonance (Suppl. Info.). The calculation has been made for
552 nm, on the blue side of the plasmon resonance, and at 566 nm, on its red side. At these
wavelengths, the scattering force is slightly lower than at resonance, while $\mathrm{F_{grad}}$
increases by at least one order of magnitude. For this reason, its oscillating character shows up
in the total force (Fig.\ref{Fig_cs_vs_wl}a, blue and red curves). Moreover,  as the polarizability
changes sign from one side to the other of the resonance, also the gradient force inverts its phase
from the blue to the red side of the resonance. Similar discussions hold for the optical forces in
front of Ag surface (Figure \ref{Fig_cs_vs_wl}b); however, in this case, the $\mathrm{F_{enz}}$ is
attractive close to the surface.

We now consider an Ag prolate spheroid as a prototypical non-spherical particle. This is chosen
with a long axis $a_1=56.8$ nm and short axes $a_2=a_3=20$ nm. As shown in Figure \ref{Fig4}e, the
particle has, in water, a long axis resonance at 560 nm and a short axis resonance at 360 nm. For
the calculation of the total optical forces we considered the case in which the spheroid has the
long axis aligned with the wave polarization, and the short semiaxis as the size parameter in Eq.
\ref{Fenz_pw1}. We obtain a further enhancement of the total optical force (tens of piconewton,
Fig.\ref{Fig4}f) which, as in the core-shell structure, is repulsive in front of ENZ surface and
attractive in front of Ag surface. In Figure \ref{Fig4}f a contour plot of the total optical force,
calculated as a function of the surface reflectivity $R$ and phase shift $\phi$, namely, in front
of all possible surfaces, is shown. We clearly see that the repulsive force can be close to 200 pN
in front of an ``ideal" ENZ surface, having the maximum reflectivity and a vanishing phase shift.

In the case of ENZ particles, we used the same $n$ and $k$ values used for the ENZ surface. We
calculated optical forces in front of glass, Ag or ENZ surfaces. The calculation has been made for
ENZ beads having radii $a$=20, 50 and 100 nm. As shown in Fig. \ref{Fig_ENZbeads_vsradius}, the
forces are about five times larger than the ones observed in dielectric bead counterparts. The
larger scattering force of ENZ particle hinders its trapping in front of glass surface, for all
radii. Moreover, the 100 nm radius ENZ particle cannot be trapped also in front of ENZ surface
(Fig. \ref{Fig_ENZbeads_vsradius}c). Results are shown in Supplementary Information.

Finally, we have studied the total optical force in case of a focused (NA=1.3) Gaussian beam,
typical of optical tweezers experiments (Section S1.2). The calculations, made for a 20 nm radius
polystyrene bead, show  that, both in front of ENZ (Fig. \ref{Fig_Gauss}a) and Ag (Fig.
\ref{Fig_Gauss}b) surfaces, the beam focusing induces a fading of the total force with the distance
$h$ (see Fig. \ref{Fig_Gauss}). The extension of the calculations for beads with larger radius
(contour plots of the total optical force in front of ENZ, Fig. \ref{Fig_Gauss}c, and Ag, Fig.
\ref{Fig_Gauss}d, surfaces) shows that the total force increases at increasing bead radius,
reaching the range of tens of femtonewton in front of ENZ and hundreds of femtonewton in front of
Ag surface. The modulation induced by the gradient force is clearly visible. It is worth noting
that when Gaussian beams are used, for a direct comparison, the beam power is reduced with respect
to the plane wave case in order to maintain the intensity at the beam focus similar to the plane
wave intensity.

\paragraph{Conclusions.} In conclusion, ENZ-based optomechanics represents a novel way to manipulate
and tailor mechanical effects of light exploiting flat surfaces. We focused our study on the
repulsive-attractive optomechanics for particles in front of an ENZ surface in realistic conditions
for a wide range of parameters (particle size and shape, ENZ surface structure, etc.) in the axial direction.
Combining the unique optical properties of ENZ metamaterials with patterning capabilities will also enable
further manipulation and control in the transverse direction towards a full dynamical engineering of ENZ-based optical forces.
Various potential applications for future study include particle sorting due to the strong dependence of ENZ-based optical forces
on the size and material composition of particles, biomolecular trapping and sensing, wavelength multiplexing of optical forces,
and chiral optical sorting, just to name a few.

\section*{Data Availability}
Data that support the findings of this study are available from the corresponding authors upon reasonable request.

\section*{Acknowledgements}
M.G.D., R.S., and O.M.M. acknowledge financial support from the agreement ASI-INAF n.2018-16-HH.0,
Project "SPACE Tweezers".  N.E. acknowledges partial support from the Vannevar Bush Faculty
Fellowship program sponsored by the Basic Research Office of the Assistant Secretary of Defense for
Research and Engineering, funded by the Office of Naval Research through Grant No.
N00014-16-1-2029. G.S. acknowledges financial support from the Ohio Third Frontier Program and the
National Science Foundation - DMR Grant No. 1708742.

\section*{Competing interests}
The authors declare no competing interests. N.E. is a strategic scientific advisor/consultant to Meta Materials, Inc.

\newpage

\centerline{\huge{Supplementary Information}}
\section*{S1 Optical forces in the dipole approximation in front of epsilon-near-zero Materials}

%\usetagform{S}

Dipole approximation (DA) is an easy and quick method to calculate optical forces on a
nanoparticle. It is valid when the size of the particle is very small compared to the wavelength of
the field \cite{Chaumet2000,AriasGonzalezJOSAA03,Jones2015}, and due to its simplicity it provides
useful results that can be compared with more complex light scattering approaches (T-matrix, DDA)
in the limit of small particles \cite{Polimeno2018}.

In DA the total optical force on a particle is usually split in a gradient force $\mathrm{F}_{\rm
grad}$ and a scattering force $\mathrm{F}_{\rm scat}$ \cite{Jones2015}:

\begin{equation}\label{Fgrad_gen}
\vec{\mathrm{{F}}}_{\rm grad}(r,z)=\frac{1}{2}\frac{n_{\rm m}}{c \varepsilon_{\rm m}}
\mathrm{Re}(\alpha) \vec{\nabla}I(r,z)
\end{equation}

\begin{equation}\label{Fscatt_gen}
\vec{\mathrm{{F}}}_{\rm scat}(r,z)=\frac{n_{\rm m} \sigma_{\rm ext}}{c}I(r,z) \hat{k}
\end{equation}

Here, $\hat{k}$ is the wave propagation direction that for an axially directed plane wave coincides
with the axial coordinate $\hat{z}$, $r$ is the radial coordinate, $c$ is the speed of light,
$\varepsilon_{\rm m}=\varepsilon_{0} n_{\rm m}^2$  is the medium permittivity, $\varepsilon_{0}$ is
the vacuum permittivity, $n_{\rm m}$ is the refractive index of the medium, $I(r,z)$ is the wave
intensity, $\alpha$ is the particle complex polarizability,

\begin{equation}\label{alpharad}
\alpha=\frac{\alpha_{0}}{1-i\frac{k^3\alpha_0}{6 \pi \varepsilon_{\rm m}}}
\end{equation}

\noindent where $\alpha_0$ is the polarizability in the static field limit (Clausius-Mossotti), and
$\sigma_{\rm ext}$ is the extinction cross-section, related to the particle absorption and
scattering \cite{AriasGonzalezJOSAA03,Jones2015}:

\begin{equation}\label{sigmaext}
\sigma_{\rm ext}=\frac{k}{\varepsilon_{\rm m}}\mathrm{Im}({\alpha})=
%\frac{k}{\varepsilon_{\rm m}} \mathrm{Im}(\alpha_0) + \frac{k^4}{6 \pi \varepsilon_{\rm m}^2} \left[ (\mathrm{Re}(\alpha_0))^2+ (\mathrm{Im}(\alpha_0))^2\right]=
\sigma_{\rm abs}+\sigma_{\rm scat}
\end{equation}

\noindent with $k=\frac{2\pi n_{\rm m}}{\lambda}$ the wave number and $\lambda$ the wavelength.

$\mathrm{{F}}_{\rm grad}$ drives the particles towards the maximum of light intensity if they have
positive polarizability; otherwise, the particles are repelled from it. On the contrary,
$\mathrm{{F}}_{\rm scat}$ always pushes the particles along field propagation direction, $\hat{k}$.
Another contribution to the total force may come from the spin-curl force \cite{Jones2015}, but
only when beams having spatial polarization gradients are used \cite{Marago2013}, which is not the
case in this work.

Recently, it has been proposed \cite{Rodriguez2014} that in front of an $\varepsilon$-near-zero
(ENZ) surface a point dipole source is subjected to a near-field repulsive force, reminding the
Meissner effect in superconductors \cite{Rodriguez2014}. In the quasistatic approximation the
near-field force is \cite{Rodriguez2014}:

\begin{equation}\label{Fenz}
\mathrm{F}_{\rm enz}(z)\approx - \sigma \frac{9}{512 \pi^4 c} \mathrm{Re}\left(
\frac{\varepsilon_{\rm s}-\varepsilon_{\rm m}}{\varepsilon_{\rm s}+\varepsilon_{\rm m}} \right)
\left( \frac{\lambda}{n_{\rm m} z} \right)^4 P_{\rm rad}
\end{equation}

\noindent where $\sigma$ is a prefactor accounting for the orientation of the dipole ($\sigma$=1,
horizontal dipole; $\sigma$=2, vertical dipole), $\varepsilon_{\rm s}$ is the complex dielectric
permittivity of the surface, $z$ is the height of the dipole above the surface and $P_{\rm rad}$ is
the radiated power of the dipole in free space.

\subsection*{S1.1 Plane wave illumination} Here, we calculate the total optical force on a finite-size particle in front of an arbitrary reflective surface in the dipole approximation. In this case, the exciting field $E_{\rm E}$ is the superposition of the incident $E_I$ and reflected  $E_R$ electromagnetic waves which produces a standing wave that, in the simplest case of plane waves travelling in the $z$ direction, can be written as:

\begin{equation}\label{Iplanewave_one}
I(z)=\frac{n_{\rm m} \varepsilon_{0}c}{2} \left| E_{\rm E}(z)\right|^2=\frac{n_{\rm m}
\varepsilon_{0}c}{2} \left| \mathrm{E}_{0} e^{-ik(z)}+ \rho \mathrm{E}_{0} e^{+ik(z)+i \phi}\right|
^2=I_0+2\rho I_0 \cos (-2kz-\phi)+\rho^2 I_0
\end{equation}

\noindent with $I_0=n_{\rm m} \varepsilon_{0}c E_0^2/2$. Note that $z$ is taken positive in the
direction of the reflected beam and $\rho$ and $\phi$ are the amplitude and phase, respectively, of
the complex reflection coefficient of the surface $r_m=\rho e^{i \phi}$, which is connected to the
surface complex refraction index  $\tilde{n}= n_{\rm s}+i k_{\rm s}$  by \cite{Born1980_6th_ed}

\begin{align}
\rho=\sqrt{\frac{(n_{\rm m} - n_{\rm s})^2 + k_{\rm s}^2}{(n_{\rm m} + n_{\rm s})^2 + k_{\rm s}^2}}
&&  \phi=\arctan \left[ \frac{-2n_{\rm m} k_{\rm s}}{n_{\rm m}^2 - n_{\rm s}^2 - k_{\rm s}^2
}\right]
\end{align}

Thus, the gradient force along the axial direction on a finite-size particle in front of a
reflective surface can be written as:

\begin{equation}\label{Fgrad_pw}
\mathrm{F_{\rm grad}}(z)=\frac{1}{2}\frac{n_{\rm m}}{c \varepsilon_{\rm m}} \mathrm{Re}(\alpha)
\frac{d I(z)}{dz}
\end{equation}

\noindent where $z=h+a$ is the axial coordinate, $h$ is the edge-to-edge distance of the particle
from the surface and $a$ is the particle radius.

The scattering force is the sum of the opposite contributions due to the incident and reflected
plane waves \cite{ZemanekOPTCOMM98b, Born1980_6th_ed, Hansen1968}:

\begin{equation}\label{Fscatt_pw}
\mathrm{F_{\rm scat}}(z)=\frac{n_{\rm m}}{c}\sigma_{\rm ext}I_0\left( \rho^2 -1  \right)
\end{equation}

\noindent where $\rho$ is related to the surface reflection coefficient
$\left|r_m\right|^2=\left|\rho e^{i\phi}\right|^2=\rho^2$.

Finally, the near-field force on the particle is:

\begin{equation}\label{Fenz_pw}
\mathrm{F_{\rm enz}}(z)\approx - \sigma \frac{9}{512 \pi^4 c} \mathrm{Re}\left(
\frac{\varepsilon_{\rm s}-\varepsilon_{\rm m}}{\varepsilon_{\rm s}+\varepsilon_{\rm m}} \right)
\left( \frac{\lambda}{n_{\rm m} z} \right)^4 \sigma_{\rm scat} I(z)
\end{equation}

\noindent where the radiated power $P_{\rm rad}$=$\sigma_{\rm ext} I(z)$ is related to the light
scattering process.

We can now add Eqs. S\ref{Fgrad_pw}-S\ref{Fenz_pw} to calculate the total optical axial force on
different types of particles in front of  dielectric, metallic or ENZ surfaces. As the dipole is
induced by a linearly polarized wave travelling ortogonally to the surface, an horizontal dipole
($\sigma$=1 in Eq. S\ref{Fenz_pw}) is used. We used an incident light intensity of about 5.6$\cdot
10^8$ W/m$^2$, corresponding to a beam power of 10 mW and a beam waist of approx. 3.5 $\mu$m, both
of which can be realized in a typical experimental configuration in our laboratories. Stable
equilibrium points for the particle dynamics are found at $z$ values in which the total optical
force vanishes with a negative slope. For small displacements from these points, the particles are
subjected to a restoring force that can be linearized as $F_{z} \approx -\kappa_z z$, with
$\kappa_z$ the trap spring constant. We consider four types of model particles: a homogeneous
dielectric (polystyrene) spherical bead, a spherical particle with parameters equivalent to an ENZ
material, a spherical core-shell particle ($\mathrm{SiO_2}$ core, Ag shell), and an Ag prolate
spheroid. The different surfaces have been considered in the calculation by means of their complex
refractive index values at 560 nm; in the ENZ case, we have chosen $n_{\rm s}\approx 0.476$ and
$k_{\rm s}\approx 0.511$ in order to obtain a  real part of complex permittivity close to zero and
an imaginary part close to 0.5. The same values have been used for the complex permittivity of the
ENZ particle.

\paragraph{Dielectric bead.} We calculated the optical forces under $\lambda$=560 nm in
water ($n_m$=1.33) on spherical polystyrene (relative permittivity 2.54) beads having radii $a$=20, 50 and 100 nm. In
this case, the Clausius-Mossotti polarizability is \cite{Bohren1998}:

\begin{equation}\label{alpha}
\alpha_0=4\pi \varepsilon_{\rm m} a^3 \left( \frac{\varepsilon_{\rm p} - \varepsilon_{\rm
m}}{\varepsilon_{\rm p}+2\varepsilon_{\rm m}} \right)
\end{equation}

In Fig. 2a of the main text, the results (short dots) obtained in DA approximation for the 20 nm
dielectric bead as a function of its distance from the different surfaces are compared with those
obtained by using more sophisticated approaches (COMSOL, cicles, and T-matrix, continuous lines). A
very good agreement is clearly observed. In all approaches, the total optical force on small
particles is modulated by the sinusoidal term in the gradient force. It is larger (in the fN range)
on more reflective surfaces and, going from Ag to ENZ, it changes phase, leading to stable traps at
different distances (arrows in Fig. 2).

The axial trap spring constants $\kappa_{\mathrm{ENZ}}$ and $\kappa_{\mathrm{Ag}}$ in front of ENZ
and Ag surfaces have been calculated by  a linear fit of the total force at the equilibrium points.
They are $\kappa_{\mathrm{ENZ}}=15$ fN/$\mu$m and $\kappa_{\mathrm{Ag}}=27$ fN/$\mu$m, that can be
compared to the trap spring constant in the axial direction obtained in a standard optical tweezers
setup, based on a single Gaussian beam. In this case,

\begin{equation}\label{Igaussian}
I(x,y,z)=I_0 \frac{w_{0}^{2}}{w(z)^{2}}\mathrm{exp}\left[-2\frac{x^2+y^2}{w(z)^2}\right]
\end{equation}

\noindent where $w_0$ is the beam waist, $w(z)=w_0\sqrt{1+\frac{(z-z_0)^2}{z_{R}^{2}}}$ is the beam
width at $z$, $z_R=\frac{n_{m}\pi w_{0}^{2}}{\lambda}$ is the Rayleigh range, $z_0$ is the position
of the beam waist and $I_0=2P/\pi w_{0}^2$ is the on-axis intensity at the waist of a beam having
total power $P$. To evaluate the beam waist, we used the Abbe criterion, $w_0=\frac{0.5
\lambda}{NA}$, with NA=1.3 the numerical aperture, as in typical single beam optical tweezers. Eqs.
S\ref{Fgrad_gen} and S\ref{Fscatt_gen} can be used to calculate the axial component of the total
force, and the corresponding $k_S$ at the equilibrium point is obtained by a linear fit. We
consider as before the particle in water and illuminated at $\lambda$=560 nm. We find, at the same
light intensity used in front of ENZ and Ag surfaces, a two order of magnitude lower
$\kappa_S=0.24$ fN/$\mu $m axial spring constant.

\begin{figure}
\includegraphics[width=\textwidth]{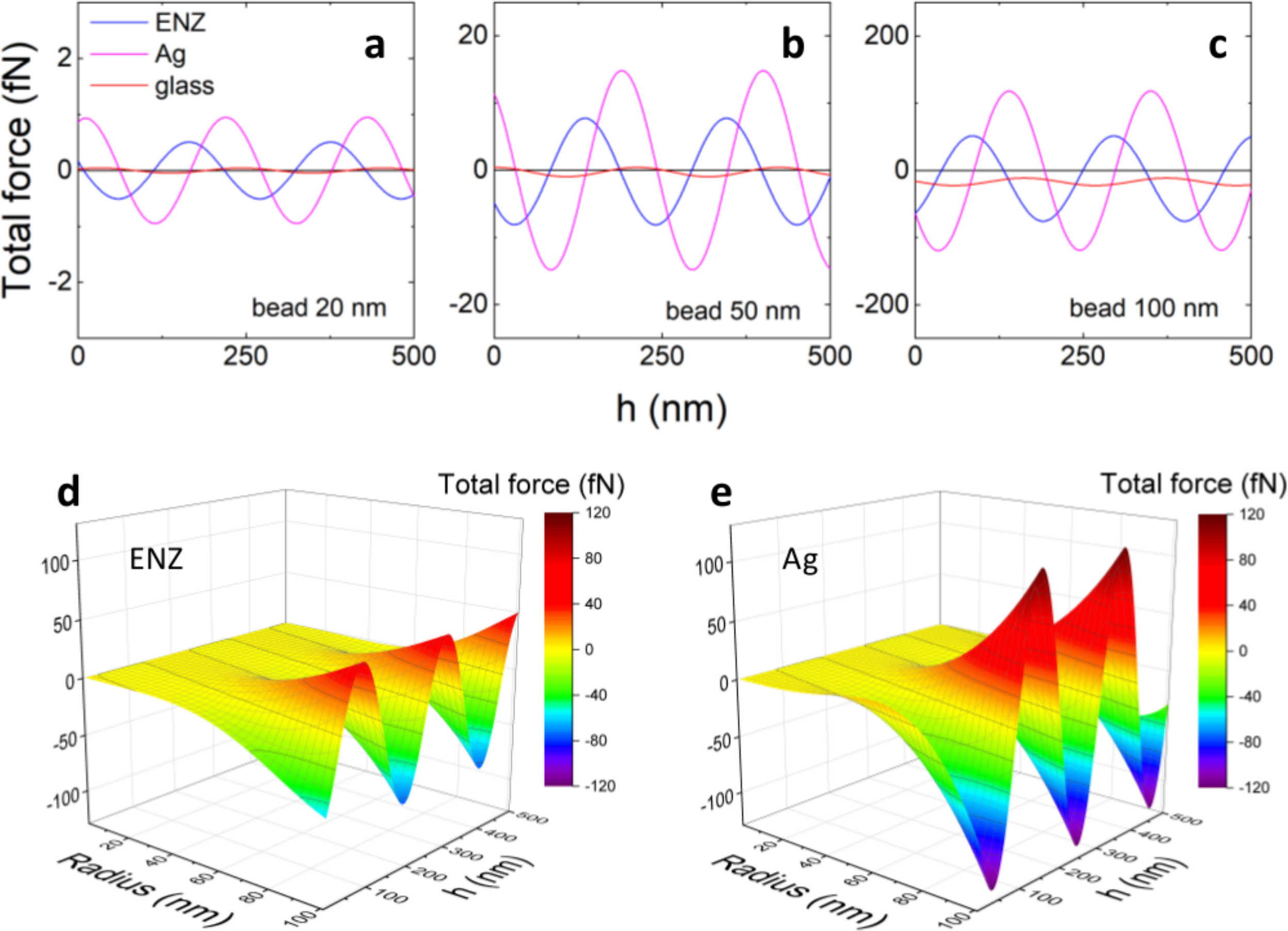}
\caption{Optical forces on a) 20 nm, b) 50 nm and c) 100 nm radius dielectric beads. The beads are
in front of ENZ (blue curve), Ag (magenta curve) or glass (red curve) surfaces. Note how for
smaller particles the dominant contribution to the optical force comes from the gradient force,
while for larger particles the greater scattering force shifts downwards the force modulation
resulting from the interference between incident and reflected field. In d) and e), 3D plots of the
total force as a function of the particle radius and of the distance from ENZ (d) and Ag (e)
surfaces.} \label{Fig_dielbeads_vsradius}
\end{figure}

In Fig. \ref{Fig_dielbeads_vsradius} the optical forces in DA on dielectric beads at increasing
bead radius (20, 50 and 100 nm) are shown. The increasing size of the particles corresponds to
larger optical forces and different trapping points. However, the 100 nm bead is not trapped in
front of glass surface, whereas it is trapped in front of ENZ and Ag surfaces, whose higher
reflectivity with respect glass surface better counteracts the scattering force due to the incoming
beam.

\begin{figure}
\includegraphics[width=\textwidth]{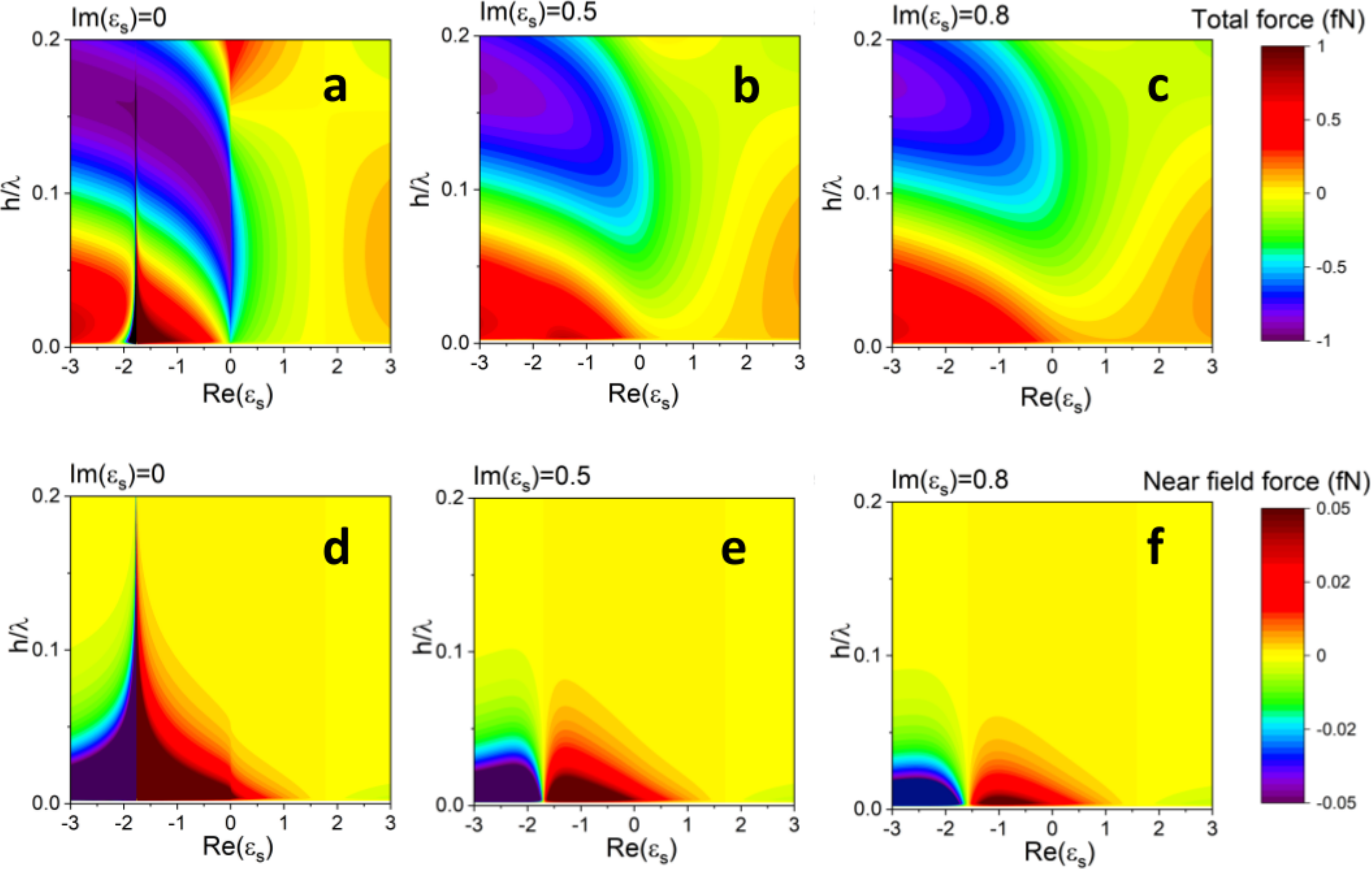}
\caption{Contour plots of the total optical force (a-c) and of the near-field force (d-f) in front
of (a,d) lossless, (b,e) medium loss, $\mathrm{Im}(\varepsilon_{\rm s})$=0.5 and (c,f) very high
loss, $\mathrm{Im}(\varepsilon_{\rm s})$=0.8 surfaces on a 20 nm dielectric bead as a function of
of the real part of the surface permittivity and of the particle normalized height $h/\lambda$
above the surface ($\lambda$=560 nm). The maximum optical force is in the fN range.}
\label{Fig_bead_all}
\end{figure}

In Fig. \ref{Fig_bead_all}, a panel summarizing the results of the calculation of the total optical
force (a-c) and of the near-field force (d-f) on a 20 nm dielectric bead in water is shown. Three
different surfaces are considered: lossless, with medium loss (Im$(\varepsilon_s)$=0.5) and with
high loss (Im$(\varepsilon_s)$=0.8). The comparison with the results obtained for a point dipole in
vacuum \cite{Rodriguez2014} shows that in this work the presence of a medium (water) broadens the
repulsive near field force region from $-1<\varepsilon_{\rm s}<1$ to $-1.77<\varepsilon_{\rm
s}<1.77$; moreover, as already observed\cite{Rodriguez2014}, even in surfaces with high loss there
is still a repulsive near field force. However, the calculation of the total optical force gives
values not higher than 1 fN, which is found only in front of lossless surfaces.

\begin{figure}
\includegraphics[width=\textwidth]{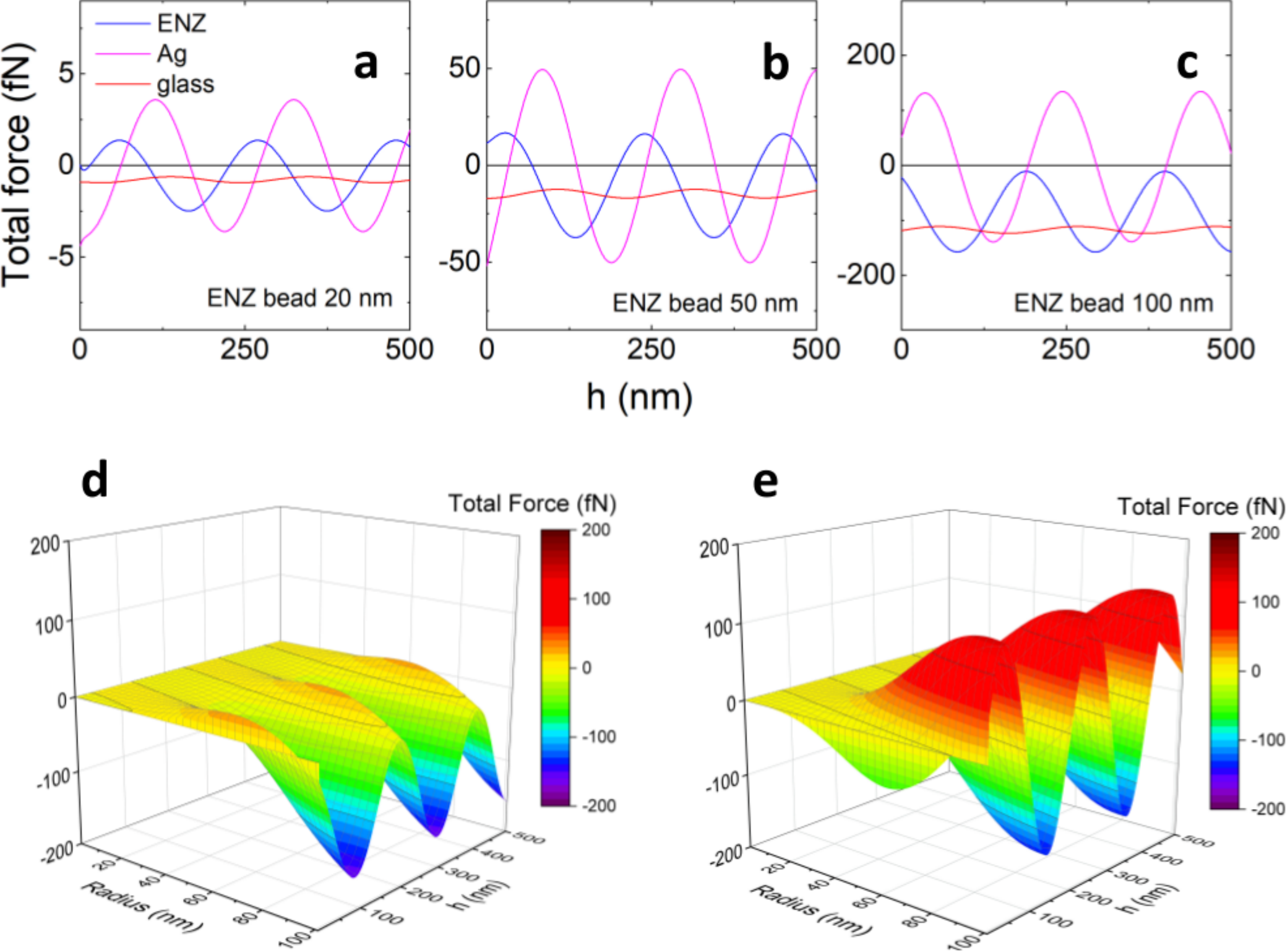}
\caption{(a-c) Total force on spherical beads made by ENZ material. The forces are shown for 20 nm
radius (a), 50 nm radius (b) and 100 nm radius (c). ENZ (blue curve), Ag (magenta) and glass (red)
surfaces are considered for the calculation. (d,e) 3D plots of the total force as a function of the
ENZ particle radius and of the distance from the ENZ (d) surface and Ag (e) surface.}
\label{Fig_ENZbeads_vsradius}
\end{figure}

\paragraph{ENZ particle.}
Optical forces on spherical beads made by ENZ material  have been calculated in front of glass, Ag
or ENZ surfaces. ENZ beads having radii $a$=20, 50 and 100 nm have been considered. As shown in
Fig. \ref{Fig_ENZbeads_vsradius}, the forces are always larger than the ones observed in dielectric
bead counterparts. The larger scattering force of ENZ particle hinders its trapping in front of
glass surface, for all radii. Moreover, the 100 nm radius ENZ particle cannot be trapped also in
front of ENZ surface (Figure \ref{Fig_ENZbeads_vsradius}c).

\begin{figure}
\includegraphics[width=\textwidth]{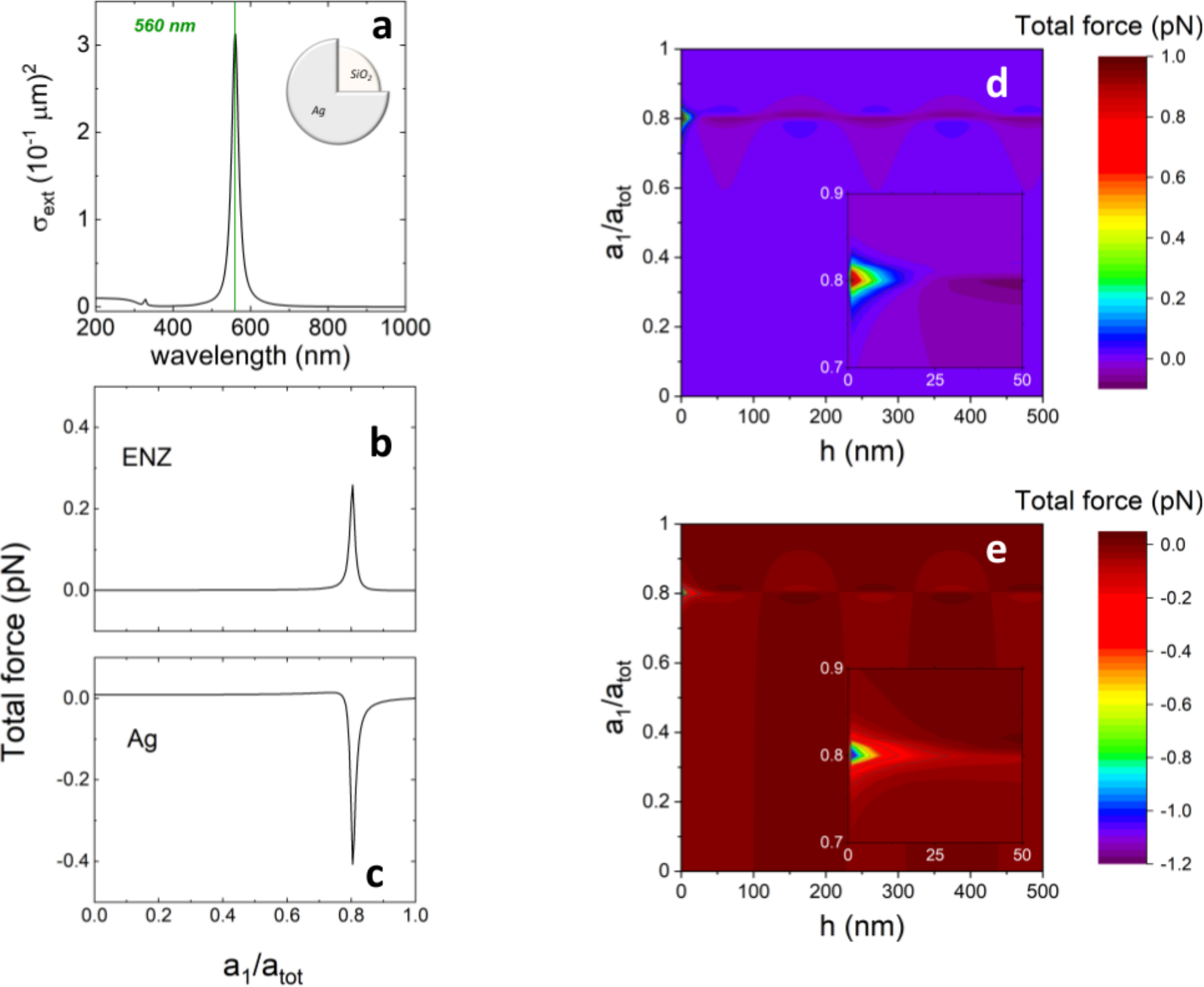}
\caption{(a) Extinction spectrum of the $\mathrm{SiO_2}$-Ag core-shell particle (total radius
$a_{tot}$=20 nm and core radius $a_1$=16.1 nm) in water. (b,c)  Total optical force of the
core-shell particle at fixed distance $h$=10 nm from ENZ (b) and Ag (c) surfaces as a function of
the $a_1$ to $a_{tot}$ ratio. (d,e) Contour plots of the optical force with respect to the $a_1$ to
$a_{tot}$ ratio and the distance $h$ from the surface. The force on the core-shell particle is in
the pN range only at short distances from the surfaces and repulsive in front of ENZ (d) while
attractive (e) in front of Ag.} \label{Fig_cs}
\end{figure}

\paragraph{Core-shell particle.}
To enhance the optical force, we used a $\mathrm{SiO_2}$-Ag core-shell particle designed to be
resonant at approximately 560 nm and having a total radius $a_{tot}$= 20 nm. The calculation of the
extinction cross-section shows (Fig. \ref{Fig_cs}a) that the resonance condition is fulfilled  if
the core-shell structure has a core radius $a_1=16.1$ nm, and the Ag shell thickness is 3.9 nm. The
particle polarizability is \cite{Bohren1998}:

\begin{equation}\label{alpha_cs}
\alpha_{cs}=4\pi a_{tot}^3 \varepsilon_{\rm m} \frac{(\varepsilon_{2}-\varepsilon_{\rm m})
(\varepsilon_{1}+2\varepsilon_{2})+f^3 (\varepsilon_{1}-\varepsilon_{2})(\varepsilon_{\rm
m}+2\varepsilon_{2})}{(\varepsilon_{2}+2\varepsilon_{\rm
m})(\varepsilon_{1}+2\varepsilon_{2})+f^3(2\varepsilon_{2}-2\varepsilon_{\rm
m})(\varepsilon_{1}-\varepsilon_{2})}
\end{equation}

In this equation, $a_{tot}$ is the core-shell total radius, $\varepsilon_{1}$ and $\varepsilon_{2}$
are the core and shell complex permittivity, respectively, and  $f=\frac{a_{1}}{a_{tot}}$ is the
ratio between the core radius $a_1$ and the total particle radius $a_{tot}$.

As shown in Figure \ref{Fig_cs}, the resonance at 560 nm enhances the optical force to the pN range
but only at very short distances from the surfaces, being repulsive in the ENZ case and attractive
in the Ag case. Otherwise, the total optical force is at the fN range.

\begin{figure}
\begin{center}
    \includegraphics[width=0.8\textwidth]{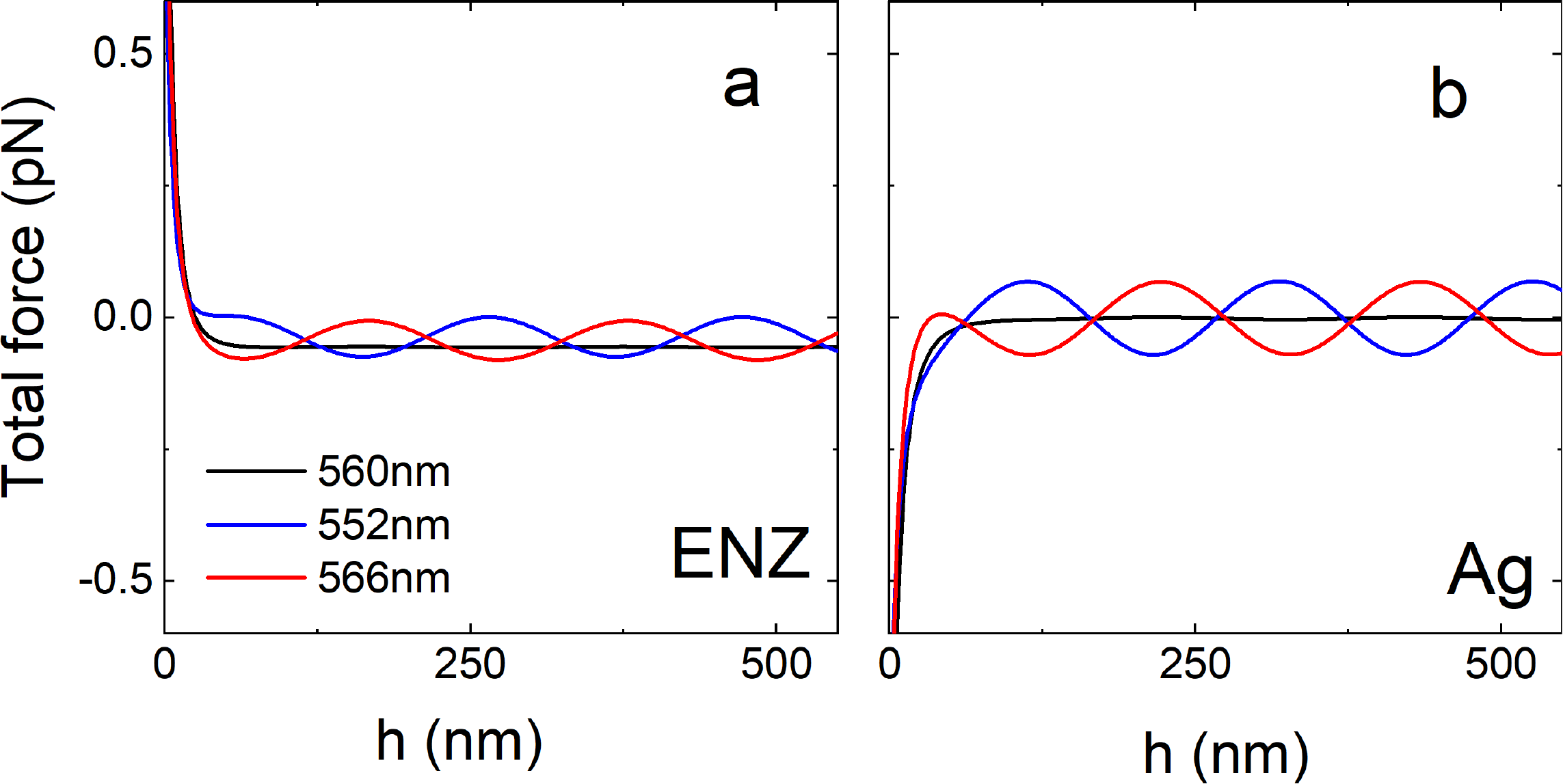}
\end{center}
\caption{Total force on a $\mathrm{SiO_2}$-Ag core-shell particle at three different wavelengths:
at resonance (560 nm, black curve), at 552 nm (blue-shifted with respect resonance, blue curve) and
at 566 nm (red-shifted with respect resonance, red curve). The total force is in the pN range close
to the surface, due to the $\mathrm{F_{enz}}$ contribution. The sinusoidal behaviour of the
gradient force is visible only out of resonance (blue and red curves), while it is negligible at
resonance, where, far from the surface, only the scattering force drives the total force. Close to
the surface, the total force is repulsive in front of ENZ and attractive in front of Ag.}
\label{Fig_cs_vs_wl}
\end{figure}

More specifically, at the resonance $\mathrm{F_{enz}}$ is in the pN range close to the ENZ surface
(from $h$=0 nm to roughly 10 nm). $\mathrm{F_{grad}}$ has an oscillating character, but its
amplitude is smaller ($\approx$ 1 fN) than $\mathrm{F_{enz}}$, due to the small real part of the
polarizability at resonance $\mathrm{Re(\alpha)=0.04 \cdot 10^{-32} \ F m^2}$. On the contrary,
$\mathrm{F_{scatt}}$ is large (tens of fN), because of the great extinction coefficient at
resonance. Thus, at 560 nm (black curve in Fig. \ref{Fig_cs_vs_wl} a), the total force is repulsive
and in the pN range close to the surface, but becomes attractive and approximately constant as the
$\mathrm{F_{enz}}$ contribution fades off.

The behaviour of the forces on the core-shell particle can also be studied for wavelengths smaller
and larger than the particle plasmon resonance. The calculation has been made for 552 nm, on the
blue side of the plasmon resonance, and at 566 nm, on its red side. At these wavelengths, the
scattering force is slightly lower than at resonance, while $\mathrm{F_{grad}}$ increase by at
least one order of magnitude. For this reason, its oscillating character now can be better noticed
in the total force (Fig.\ref{Fig_cs_vs_wl} a, blue and red curves). Moreover,  as the
polarizability changes sign from one side to the other of the resonance, also the gradient force is
``out of phase'' going from the blue to the red side of the resonance. Similar discussions hold
also for the calculation of forces in front of Ag surface (Figure \ref{Fig_cs_vs_wl} b); however,
in this case, the $\mathrm{F_{enz}}$ is attractive close to the surface.

\paragraph{Ag prolate spheroid.}
We choose an Ag prolate spheroid having long axis $a_1=56.8$ nm and short axes $a_2=a_3=20$ nm. The
particle polarizability is

\begin{equation}\label{alpha1}
\alpha_{i}=\frac{4}{3}\pi a_1 a_2 a_3 \varepsilon_{\rm m} \frac{\varepsilon_{\rm
p}-\varepsilon_{\rm m}}{\varepsilon_{\rm m}+L_i (\varepsilon_{\rm p}-\varepsilon_{\rm m})}
\end{equation}

\noindent In this equation,  $\varepsilon_{\rm p}$ is the particle permittivity and $L_i$ is a
geometric factor relative to the spheroid axis $a_i$. In case of a prolate spheroid, $L_1$ is
\begin{align}\label{L1}
L_1=\frac{1-e^2}{e^2}\left( -1+\frac{1}{2e}\mathrm{ln}\frac{1+e}{1-e} \right) &&
e^2=1-\frac{a_{2}^{2}}{a_{1}^{2}}
\end{align}

\noindent and $L_2=L_3=\frac{1}{2}(1-L_1)$.

\begin{figure}
\includegraphics[width=\textwidth]{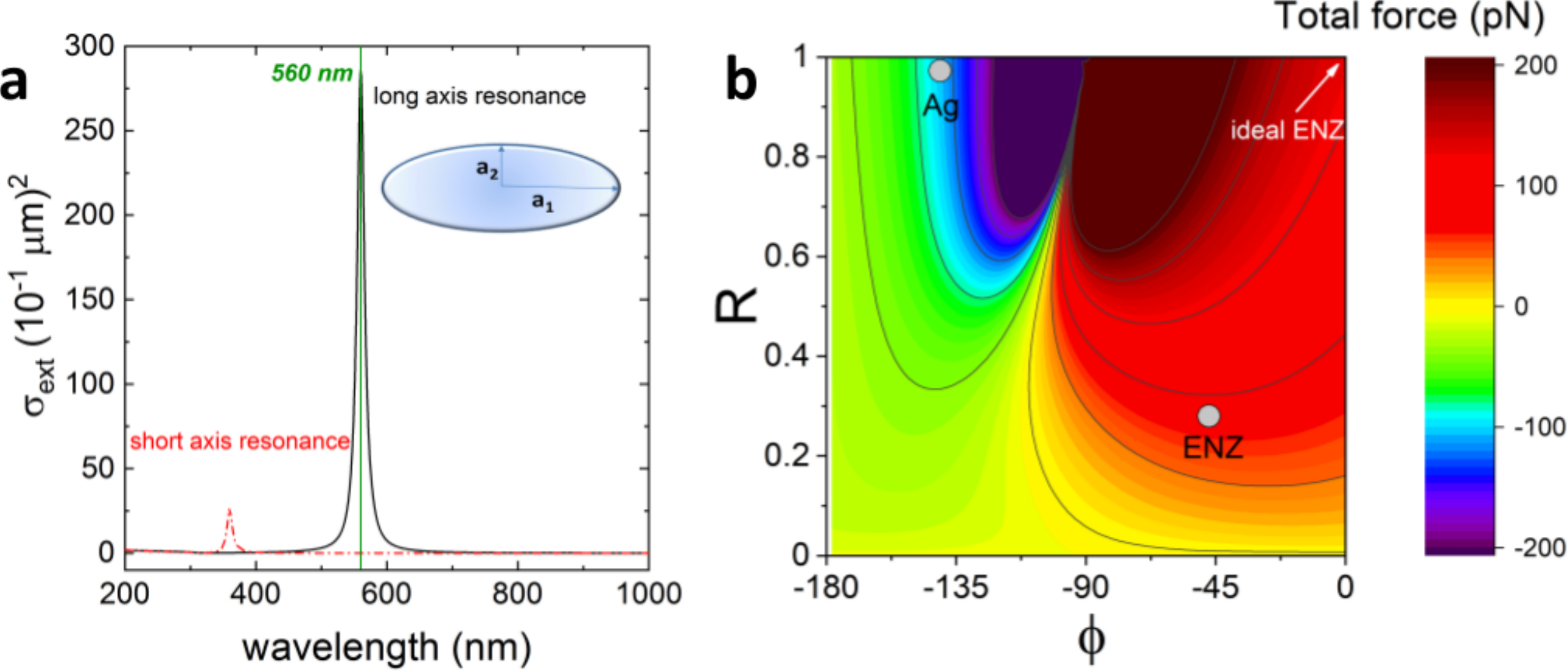}
\caption{(a) Extinction spectra of Ag prolate ellipsoid in water oriented with the long axis
parallel to the field (black solid line) and oriented with the short axis parallel to the field
(red dashed line). The resonances relative to the long and short axes are indicated. (b) Contour
plot of the total optical force on the Ag ellipsoid at $h$=10 nm distance from a surface as a
function of the surface reflectivity R and phase shift $\phi$. In the calculation, the spheroid is
aligned with the long axis in the direction of the wave polarization. The  ENZ and Ag surfaces used
for the calculation of the optical forces in DA approximation are shown. The optical force is in
the order of tens of pN in front of ENZ (repulsive) and Ag (attractive). The force can be close to
200 pN if the spheroid is in front of an ideal ENZ surface, having R=1 and $\phi$=0.   }
\label{Fig_Ag_ell}
\end{figure}

As shown in Figure \ref{Fig_Ag_ell}a, the particle has, in water, a long axis resonance at 560 nm
and a short axis resonance at 360 nm. For the calculation of the total optical force we considered
the case in which the spheroid has the long axis aligned with the wave polarization, so to use
$\alpha_1$ for the polarizability in Eqs. S\ref{Fgrad_pw} and S\ref{sigmaext}, and the short
semiaxis as the size parameter in Eq. S\ref{Fenz_pw}. We obtain a further enhancement of the total
optical force (tens of pN) which, as in the core-shell structure, is repulsive in front of ENZ
surface and attractive in front of Ag surface. In Figure \ref{Fig_Ag_ell}b a contour plot of the
total optical force, calculated as a function of the surface reflectivity $R$ and phase shift
$\phi$, namely, in front of all possible surfaces, is shown. We easily see that the repulsive force
can be close to 200 pN in front of an ``ideal" ENZ surface, having the maximum reflectivity and a
vanishing phase shift.

\subsection*{S1.2 Gaussian beams}
In optical tweezers, light beams are tightly focused in order to increase $\mathrm{F_{\rm grad}}$
with respect to $\mathrm{F_{\rm scat}}$. We can introduce this condition in our calculations by
using Gaussian beams \cite{ZemanekOPTCOMM98b} instead of plane waves:

\begin{equation}\label{Einc}
\mathrm{E}_{\rm
I}(z,r)=\mathrm{E}_{0}\frac{w_0}{w_i(z)}\mathrm{exp}\left(-\frac{r^2}{w_{i}^2(z)}\right)\mathrm{exp}\left[
-ik(z+z_0)+\frac{i}{2}\frac{kr^2}{R_i}+i\arctan \left(\frac{z+z_0}{z_R}\right)\right]
\end{equation}

\begin{equation}\label{Erefl}
\mathrm{E}_{\rm
R}(z,r)=\mathrm{E}_{0}\rho\frac{w_0}{w_r(z)}\mathrm{exp}\left(-\frac{r^2}{w_{r}^2(z)}\right)\mathrm{exp}\left[
+ik(z-z_0)+\frac{i}{2}\frac{kr^2}{R_r}-i\arctan \left(\frac{z-z_0}{z_R}\right)+i\phi \right]
\end{equation}

Here, $w_0$ is the beam waist,  $z_R=\frac{n_{m}\pi w_{0}^{2}}{\lambda}$ is the Rayleigh range,
$z_0$ is the position of the beam waist, $R_i$ and $R_r$ are the wave curvature radii of the
incident and reflected wave, respectively, and $w_i(z)$ and $w_r(z)$ are the beam widths at $z$
distance:

\begin{align}
w_i (z)=w_0\sqrt{1+\frac{(z+z_0)^2}{z_{R}^2}} && w_r(z)=w_0\sqrt{1+\frac{(z-z_0)^2}{z_{R}^2}}
\end{align}

For the sake of simplicity, we restrict ourselves to the calculation of the optical force along the
beam propagation axis. The light intensity distribution $I(z)$ is \cite{ZemanekOPTCOMM98b}:

\begin{equation}\label{Intensity}
I(z)=I_0 \frac{w_{0}^{2}}{w_{i}^{2}(z)}+2\rho I_0\frac{w_{0}^{2}}{w_{i}(z)w_{r}(z)}\cos
(\psi(z))+\rho^2 I_0 \frac{w_{0}^{2}}{w_{r}^{2}(z)}
\end{equation}
Here, $I_0=2P/\pi w_{0}^2$ is the on-axis intensity at the waist of a beam having total power $P$
and

\begin{equation}\label{totalphase}
\psi(z)=-2kz + \arctan (\frac{z+z_0}{z_R})+\arctan (\frac{z-z_0}{z_R})-\phi
\end{equation}

\noindent is a factor due to the phase shift of the beam on reflection from the surface. Thus,
optical forces are calculated from Eqs. S\ref{Fgrad_gen}, S\ref{Fscatt_gen} and S\ref{Fenz}.

\begin{figure}
\includegraphics[width=\textwidth]{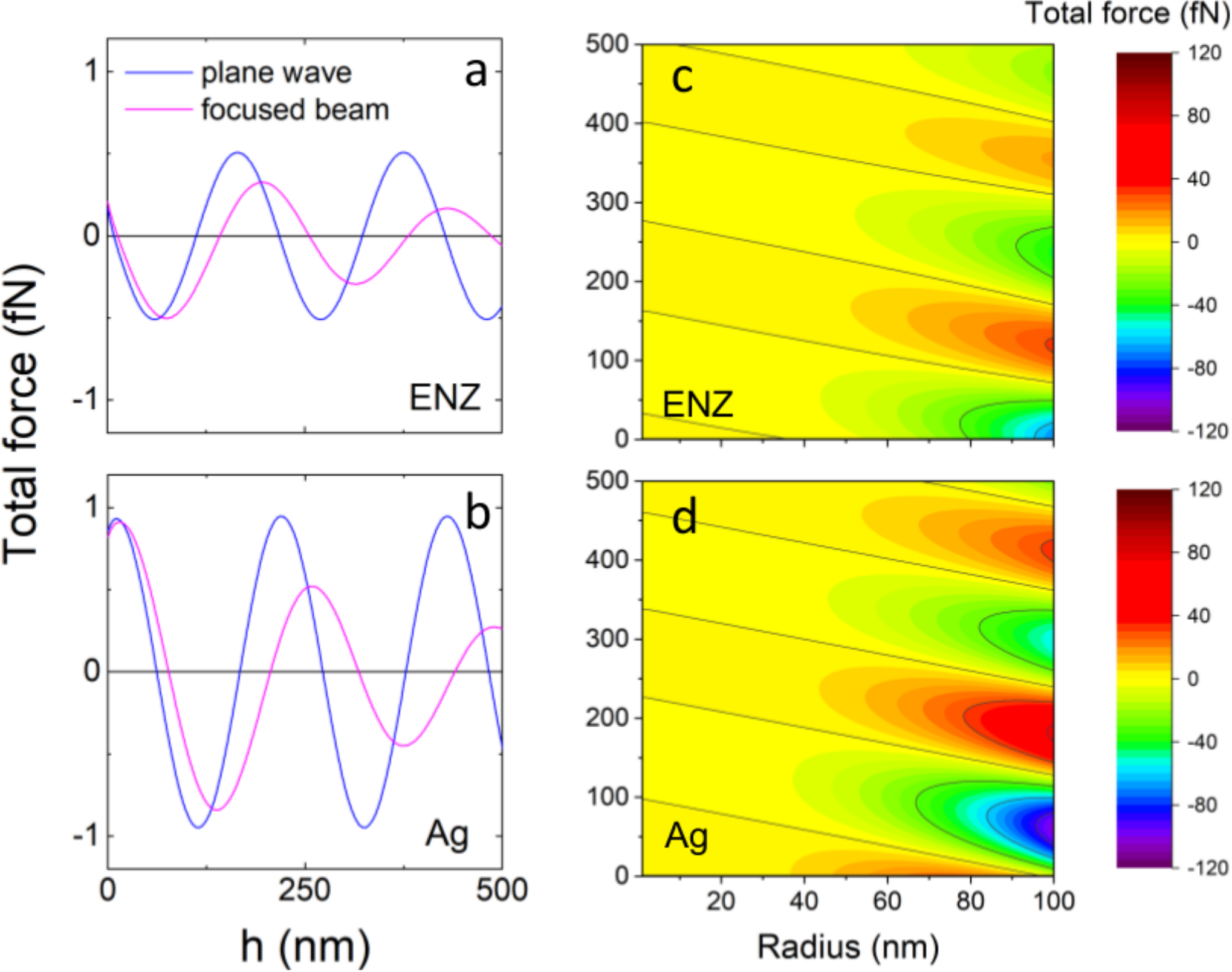}
\caption{(a,b) Total optical force on a 20nm dielectric bead under plane wave (blue curves) and
focused Gaussian beam (red curves), in front of ENZ (a) and Ag (b) surfaces, as a function of the
distance h from the surface. The focusing induces a fading of the total force with $h$. Note that
the Gaussian beam power is reduced with respect to the plane wave case in order to maintain fixed
the intensity at the beam focus. (c,d) Contour plots of the total optical force on a dielectric
bead under focused Gaussian beam illumination in front of ENZ (c) and Ag (d) surfaces, as a
function of the distance $h$ from the surface and of the bead radius. The modulation due to the
sinusoidal term in the gradient force is clearly visible. The total force increases at increasing
bead radius, reaching the range of tens of fN in front of ENZ and hundreds of fN in front of Ag
surface. } \label{Fig_Gauss}
\end{figure}

As above, our calculations consider a particle in water ($n_m$=1.33) and under illumination at
$\lambda$=560 nm; moreover, to evaluate the beam waist, we use the Abbe criterion
$w_0=0.5\frac{\lambda}{NA}$,  where the numerical aperture (NA) of the beam is NA=1.3, as in
typical optical trapping experiments. The comparison between the results obtained with both plane
wave and Gaussian beam on a small dielectric bead (radius 20 nm) are shown in Fig. \ref{Fig_Gauss}.
It is worth noting that when Gaussian beams are used, the beam power is reduced with respect to the
plane wave case in order to maintain fixed the intensity at the beam focus.

\section*{S2 Electromagnetic scattering theory and T-matrix formalism in front of epsilon-near-zero materials}
We use two different modeling approaches based on the T-matrix formalism and on finite elements
methods (COMSOL), respectively. In particular, electromagnetic scattering from particles near to or
deposited on a plane surface that separates two homogeneous media of different optical properties
in the T-matrix formalism\cite{Borghese2007book,JOSA95,JOSA99,AO99} can give account on the role of
the different multipoles in the particle-surface interaction. Indeed, the presence of the surface
can have a striking effect on the scattering pattern from the particles since the exciting field
does not coincide with the incident plane wave and the observed field does not coincide with the
field scattered by the particle. The field that illuminates the particles is partly or totally
reflected by the surface and the reflected field contributes both to the exciting and to the
observed field. Moreover, the field scattered by the particles is reflected by the interface and
thus contributes to the exciting field. In other words there are multiple scattering processes
between the particles and the interface. As a result, the field in the accessible half-space
includes the incident field $\mathbf{E}_{\rm I}$, the reflected field $\mathbf{E}_{\rm R}$, the
scattered field $\mathbf{E}_{\rm S}$ and, finally, the field $\mathbf{E}_{\rm SR}$ that after
scattering by the particles is reflected by the surface.

The mathematical difficulties that are met in calculating the scattering pattern are due to the
need that the field in the accessible half-space satisfy the boundary conditions both across the
(closed) surface of the particles and across the (infinite) interface. In other words, even by
assuming that we are able to impose the boundary conditions across the surface of the particle, the
problem still remains of imposing the boundary conditions across the
interface\cite{JOSA95,JOSA99,AO99}. It is possible to define the transition matrix for particles in
the presence of the interface that is the starting point to calculate optical forces and torques
either by direct integration of the Maxwell stress tensor or by exploiting the general expressions
of optical force and torque in terms of multiple
expansion\cite{Saija2005,Borghese2006,Borghese2007}.

\begin{figure}
\centerline{\includegraphics[width=0.7\textwidth]{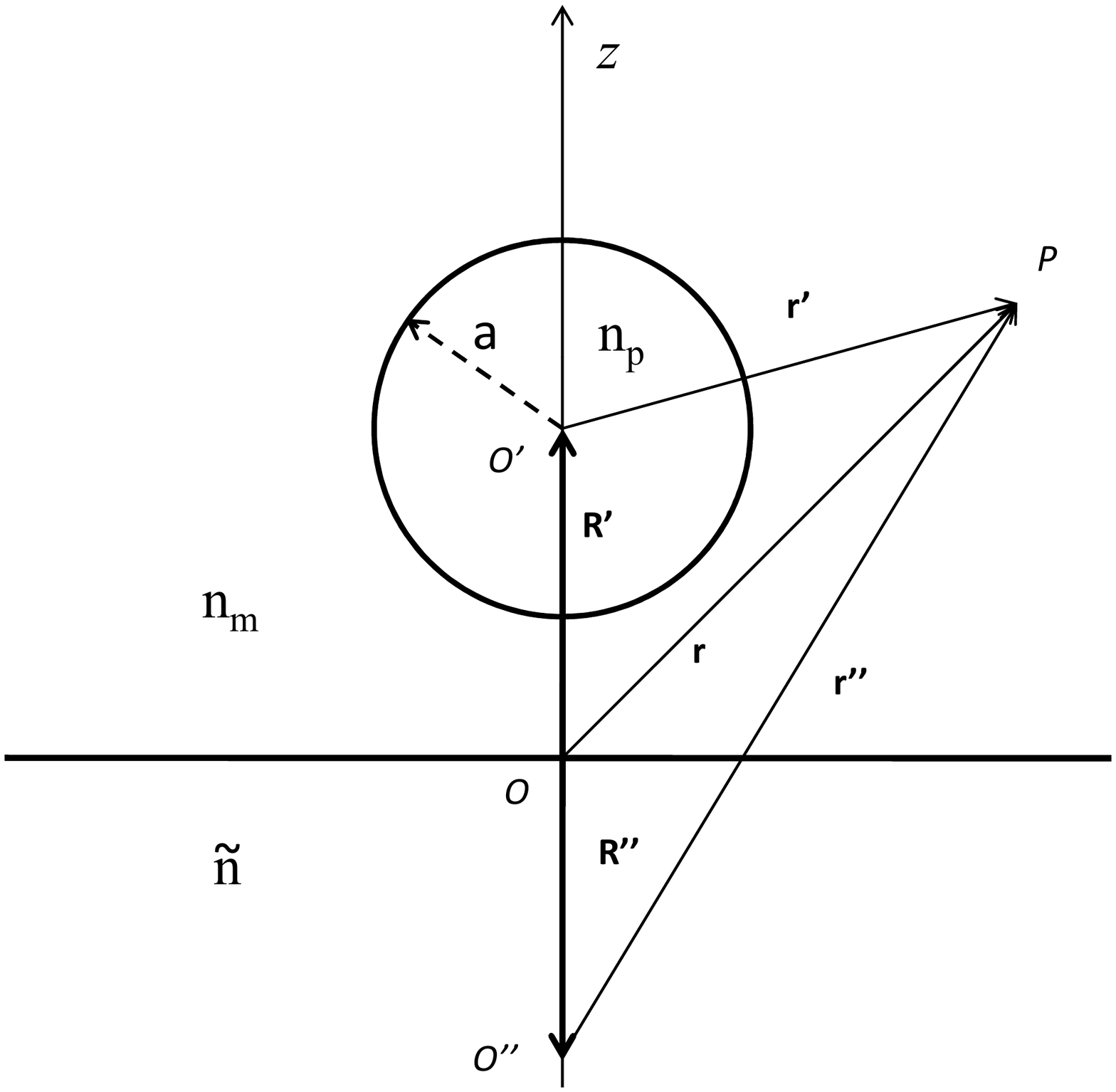}} \caption{Geometry adopted for
electromagnetic scattering from a sphere in the vicinity of a surface.}\label{Fig_geom}
\end{figure}

\paragraph{Incident and Reflected Fields.} The reflection of a
plane wave on a plane surface can be dealt with in general terms, \textit{i.e.}, without specifying
whether the medium that fills the not accessible half-space is a dielectric or a metal. This
information can, indeed, be supplied at the end of the algebraic manipulations. Let us thus assume
that the interface is the plane $z=0$ of a Cartesian frame of reference and that the half-space
$z>0$, which we take as the accessible half-space, is filled by a homogeneous  medium of (real)
refractive index $n_m$. The half-space $z<0$ is assumed to be filled by a homogeneous medium with
(possibly complex) refractive index $\tilde{n}$. Figure \ref{Fig_geom} shows the adopted geometry.
The plane wave field
\begin{equation}
\mathbf{E}_{\rm I}=E_{0}\hat{\mathbf{e}}_{\rm I} \exp(i \mathbf{k}_{\rm I}\cdot \mathbf{r})\;,
\end{equation}
which propagates within the accessible half-space, is reflected by the interface into the plane
wave
\begin{equation}
\mathbf{E}_{\rm R}=E'_{0}\mathbf{\hat{e}}_{\rm R}\exp(i \mathbf{ k}_{\rm R} \cdot \mathbf{ r})\;,
\end{equation}
where $\mathbf{ k}_{\rm I}=k'\mathbf{\hat{k}}_{\rm I}$ and $\mathbf{ k}_{\rm
R}=k'\mathbf{\hat{k}}_{\rm R}$ are the propagation vectors of the incident and of the reflected
wave, respectively, $k'=n_m k$ and $\mathbf{\hat{e}}_{\rm I}$ and $\mathbf{\hat{e}}_{\rm R}$ are
the respective unit polarization vectors. The polarization is analyzed with respect to the two
pairs of unit vectors $\mathbf{\hat{u}}_{\rm I\eta}$ and $\mathbf{\hat{u}}_{\rm R\eta}$ that are
parallel ($\eta=1$) and perpendicular ($\eta=2$) to the plane of incidence that, as usual, is
defined as the plane that contains $\mathbf{ k}_{\rm I}$, $\mathbf{ k}_{\rm R}$ and the $z$ axis.
Our choice of the orientation is defined by the equations
\begin{equation}
\mathbf{\hat{u}}_{\rm I1}\times\mathbf{\hat{u}}_{\rm I2}=\mathbf{\hat{k}}_{\rm I}\;,\qquad
\mathbf{\hat{u}}_{\rm R1}\times\mathbf{\hat{u}}_{\rm R2}=\mathbf{\hat{k}}_{\rm R}\;,
\end{equation}
with $\mathbf{\hat{u}}_{\rm I2}\equiv\mathbf{\hat{u}}_{\rm R2}$. In terms of the projections on the
polarization basis, the incident and the reflected field can be written
\begin{equation}
\mathbf{E}_{\rm I}=E_{0}\sum_{\eta}(\mathbf{\hat{e}}_{\rm I}\cdot \mathbf{\hat{u}}_{\rm
I\eta})\mathbf{\hat{u}}_{\rm I\eta}\exp(i \mathbf{ k}_{\rm I} \cdot \mathbf{ r})\;,
\end{equation}
and
\begin{equation}
\mathbf{E}_{\rm R}=E'_{0}\sum_{\eta}(\mathbf{\hat{e}}_{\rm R}\cdot \mathbf{\hat{u}}_{\rm
R\eta})\mathbf{\hat{u}}_{\rm R\eta}\exp(i \mathbf{ k}_{\rm R} \cdot \mathbf{ r})\;.
\end{equation}
In the preceding equations the incident field $\mathbf{E}_{\rm I}$ and the reflected field
$\mathbf{E}_{\rm R}$ are decomposed into their components parallel and orthogonal to the plane of
incidence and can be referred to each other by means of the Fresnel coefficients $F_{\eta}$ for the
reflection of a plane wave with polarization along $\mathbf{\hat{u}}_{\eta}$.

%%daqui
Requiring the continuity of the normal and tangential components of the fields, the reflection
condition\cite{Jackson_new} yields the equation
\begin{equation}\label{6.1}
E'_{0}(\mathbf{\hat{e}}_{\rm R}\cdot\mathbf{\hat{u}}_{\rm R\eta})=E_{0}F_{\eta}(\vartheta_{\rm
I})(\mathbf{\hat{e}}_{\rm I}\cdot\mathbf{\hat{u}}_{\rm I\eta})\;,
\end{equation}
where  the Fresnel coefficients are defined as
\begin{equation}
F_{1}(\vartheta_{\rm I})=\frac{\bar{n}^{2}\cos\vartheta_{\rm
I}-\bigl[(\bar{n}^{2}-1)+\cos^{2}\vartheta_{\rm I}]^{1/2}} {\bar{n}^{2}\cos\vartheta_{\rm
I}+\bigl[(\bar{n}^{2}-1)+\cos^{2}\vartheta_{\rm I}\bigr]^{1/2}}\;, \quad F_{2}(\vartheta_{\rm I})=
\frac{\cos\vartheta_{\rm I}-\bigl[(\bar{n}^{2}-1)+\cos^{2}\vartheta_{\rm
I}\bigr]^{1/2}}{\cos\vartheta_{\rm I}+\bigl[(\bar{n}^{2}-1)+\cos^{2}\vartheta_{\rm
I}\bigr]^{1/2}}\;,
\end{equation}
in which $\vartheta_{\rm I}$ is the angle between $\mathbf{\hat{k}}_{\rm I}$ and the $z$ axis,
$\bar{n}=\tilde{n}/n_m$. The reflected wave can be rewritten as
\begin{equation}
\mathbf{E}_{\rm R}=E_{0}\sum_{\eta}F_{\eta}(\vartheta_{\rm I}) (\mathbf{\hat{e}}_{\rm
I}\cdot\mathbf{\hat{u}}_{\rm I\eta})\mathbf{\hat{u}}_{\rm R\eta}\exp(i \mathbf{ k}_{\rm R}\cdot
\mathbf{ r})\;.
\end{equation}
The incident and the reflected field, solutions of Helmholtz equation in accessible free space, can
be expanded in terms of a series of spherical vector multipole fields centered on a suitable common
origin, $O$. To ensure the regularity of the fields at the origin, we choose J-multipole fields
defined in terms of spherical radial Bessel functions ${j}_{\rm l}(k'r)$
\cite{Jackson_new,Borghese2007book}. The result is
\begin{align*}
\mathbf{E}_{\rm I}=&\sum_{\eta}E_{0\eta}\sum_{plm}\mathbf{ J}^{(p)}_{lm}(\mathbf{
r},k')W^{(p)}_{{\rm I}\eta lm}\;,\\
\mathbf{E}_{\rm R}=&\sum_{\eta}E_{0\eta}F_{\eta}(\vartheta_{\rm I}) \sum_{plm}\mathbf{
J}^{(p)}_{lm}(\mathbf{ r},k')W^{(p)}_{{\rm R}\eta lm}\;,
\end{align*}
where the incident and reflected amplitudes are respectively:
\begin{equation}\label{6.2'}
W^{(p)}_{{\rm I}\eta lm}= W^{(p)}_{lm}(\mathbf{\hat{u}}_{\rm I\eta},\mathbf{\hat{k}}_{\rm I})\;
\end{equation}
\begin{equation}\label{6.2'}
W^{(p)}_{{\rm R}\eta lm}= W^{(p)}_{lm}(\mathbf{\hat{u}}_{\rm R\eta},\mathbf{\hat{k}}_{\rm R}).\;
\end{equation}
Because of the reflection condition due to the presence of the surface, the incident and reflected
amplitudes are not mutually independent. Infact, as the polar angles of $\mathbf{\hat{u}}_{\rm R1}$
and $\mathbf{\hat{u}}_{\rm R2}$ are
\begin{equation}
\vartheta_{\rm R1}=\vartheta_{\rm I}+\frac{\pi}{2},\quad\varphi_{\rm R1}=\varphi_{\rm
I}+\pi\;,\quad\text{and}\quad \vartheta_{\rm R2}=\frac{\pi}{2},\quad\varphi_{\rm R2}=\varphi_{\rm
I}+\frac{\pi}{2}\;,
\end{equation}
we get
\begin{equation}\label{6.3}
W^{(p)}_{{\rm R}\eta lm}=(-)^{\eta+p+l+m}W^{(p)}_{{\rm I}\eta lm}\;.
\end{equation}
In this way the amplitudes of the reflected field never need to be explicitly considered, and
conveniently we can define the exciting field as the superposition of incident and reflected fields
\begin{equation}\label{6.4}
\mathbf{E}_{\rm E}=\mathbf{E}_{\rm I}+\mathbf{E}_{\rm R}\;.
\end{equation}

As a consequence the multipole expansion of $\mathbf{E}_{\rm E}$ can be written in a more compact
form as
\begin{equation}\label{6.5}
\mathbf{E}_{\rm E\eta}=E_{0}\sum_{plm}\mathbf{J}^{(p)}_{lm}(\mathbf{r},k') W^{(p)}_{{\rm E}\eta lm}
\end{equation}
with
\begin{equation}\label{6.6}
W^{(p)}_{{\rm E}\eta lm}=[1+F_{\eta}(\vartheta_{\rm I})(-)^{\eta+p+l+m}] W^{(p)}_{{\rm I}\eta
lm}\;.
\end{equation}

\paragraph{Scattering from a Sphere on a Plane Surface.}
We assume that a spherical scatterer lies entirely within the accessible half-space and is
illuminated by a plane wave. Outside the scatterer the total field is
\begin{equation}\label{6.27}
\mathbf{ E}_{\rm Ext}=\mathbf{ E}_{\rm E}+\mathbf{ E}_{\rm S}+\mathbf{ E}_{\rm SR}\;,
\end{equation}
where $\mathbf{ E}_{\rm E}=\mathbf{ E}_{\rm I}+\mathbf{E}_{\rm R}$ is the same as we would have if
no particle were present. $\mathbf{E}_{\rm S}$, the field scattered by the sphere,  and
$\mathbf{E}_{\rm SR}$, the field that after scattering by the particle is reflected by the surface,
are related to each other by the reflection condition. Their superposition represent the observed
scattered field that we indicate with $\mathbf{E}_{\rm Obs}$.

The field that is scattered by a sphere that lies entirely in the accessible half-space can be
expanded in a series of vector H-multipole fields that satisfy the radiation condition at infinity.
The multipole fields are defined in terms of spherical radial Hankel
functions\cite{Borghese2007book}  ${h}_{\rm l}(k'r)$ Choosing for the scattered field the origin
$O'$ within the particle, we obtain
\begin{equation}\label{6.28}
\mathbf{E}_{\rm S\eta}=E_{0\eta}\sum_{plm}\mathbf{ H}^{(p)}_{S,lm}(\mathbf{
r}',k')\mathcal{A}^{(p)}_{\eta lm}\;.
\end{equation}
where the unknown amplitudes $\mathcal{A}$ can be determined by applying the boundary conditions at
the particle's surface. The asymptotic expression of $\mathbf{E}_{\rm S\eta}$ can be written easily
as follows
\begin{equation}\label{6.30}  %(6.37)
\mathbf{H}^{(p)}_{{\rm F}lm}=-\frac{\rm i}{4\pi k'}\frac{{\rm e}^{{\rm i}
k'r'}}{r}\sum_{\eta'}\mathbf{\hat{u}}_{{\rm S}\eta'}W^{(p)\ast}_{{\rm S}\eta'lm}\;.
\end{equation}
These are the multipole fields that enter in the definition of scattering amplitude of the system.

The scattered field $\mathbf{E}_{\rm S\eta}$ impinges on the plane surface and, by reflection
yields a reflected-scattered field in the vicinity of the surface of the particle. Thanks to the
reflection rule of $\mathbf{H}$-vector multipole fields\cite{fucile1997general}, that proves the
fields are given by a superposition of J-multipole vector fields with origin at  $O'$, we get:
\begin{equation}\label{6.29}
\mathbf{ E}_{\rm SR\eta}=E_{0\eta}\sum_{plm}\sum_{p'l'} \mathbf{ J}^{(p)}_{lm}(\mathbf{ r}',k')
\mathcal{F}^{(pp')}_{ll';m}\mathcal{A}^{(p')}_{\eta l'm}\;,
\end{equation}
The quantities $\mathcal{F}^{(pp')}_{ll';m}$  can be understood as the elements of a diagonal
matrix F that effects the reflection of the H-multipole fields on the plane interface giving the
formal solution to the problem. Assuming that the scattering particle is a homogeneous sphere with
(possibly complex) refractive index $n_{\rm p}$ and radius $a$, also the field regular at $O'$
within the sphere can be expanded in the form
\begin{equation}
\mathbf{ E}_{\rm T\eta}=E_{0\eta}\sum_{plm} \mathbf{ J}^{(p)}_{lm}(\mathbf{ r}',k_{\rm
p})\mathcal{C}^{(p)}_{\eta lm}\;.
\end{equation}
The boundary conditions at the surface of the sphere between the external total field,
$\mathbf{E}_{\rm E}+\mathbf{ E}_{\rm S}+\mathbf{ E}_{\rm SR}\ $, and the field within the
scatterer, $\mathbf{ E}_{\rm T}$, can be applied provided that the exciting field $\mathbf{ E}_{\rm
E}$ is referred to the center of the sphere, $O'$. This can be done resorting to the appropriate
phase factors: $\exp(i \mathbf{ k}_{\rm I}\cdot \mathbf{ R}')$ and $\exp(i \mathbf{ k}_{\rm R}\cdot
\mathbf{ R}')$. For each $p$, $l$, and $m$, we obtain four equations among which the amplitudes of
the internal field $C$ can be easily eliminated. As a result, we get, for each $m$, a system of
linear nonhomogeneous equations for the amplitudes $\mathcal{A}_{\eta lm}^{(p)}$, namely
\begin{equation}\label{6.30}
\sum_{p'l'}\mathcal{M}^{(pp')}_{ll';m}\mathcal{A}_{\eta l'm}^{(p')}= -\mathcal{W}^{(p)}_{\eta
lm}\;,
\end{equation}
where
\begin{equation}\label{6.31}
\mathcal{M}^{(pp')}_{ll';m}=\bigl({R}^{(p)}_{l}\bigr)^{-1}
\delta_{pp'}\delta_{ll'}+\mathcal{F}^{(pp')}_{ll';m}\;,
\end{equation}
and
\begin{equation}\label{6.32}
\mathcal{W}^{(p)}_{\eta lm}=\exp(i \mathbf{ k}_{\rm I}\cdot \mathbf{ R}') W^{(p)}_{{\rm I}\eta lm}
+\exp(i \mathbf{ k}_{\rm R}\cdot \mathbf{ R}')F_{\eta} W^{(p)}_{{\rm R}\eta lm}\;.
\end{equation}
The quantities ${R}^{(1)}_l$ and ${R}^{(2)}_l$ coincide with the Mie coefficients $b_l$ and $a_l$,
respectively, for a homogeneous sphere of refractive index $n_{\rm p}$ embedded into a homogeneous
medium of refractive index $n_{\rm m}$. We remark that our theory can easily deal also with sphere
sustaining longitudinal waves (plasmonic particles) or with radially nonhomogeneous
spheres\cite{Borghese2007book}.

Once the amplitudes $\mathcal{A}^{(p)}_{\eta lm}$ of $\mathbf{ E}_{\rm S\eta}$ have been calculated
by solving (\ref{6.30}), the reflected-scattered field $\mathbf{ E}_{\rm SR\eta}$ is also
determined by (\ref{6.29}).  A brief comment on the expression of the reflected-scattered field is
in order. $\mathbf{E}_{\rm SR\eta}$ is valid only in the vicinity of the surface of the sphere as
it includes multipole fields that do not satisfy the radiation condition at infinity, for this
reason to get the reflected-scattered field that would be observed by an optical instrument in the
far zone it is necessary to cast  $\mathbf{E}_{\rm SR\eta}$ in its asymptotic form. At any point of
the accessible half-space,  $\mathbf{ E}_{\rm FSR\eta}$ is given by the equation
\cite{wriedt1998light}
\begin{equation}\label{6.33}
\mathbf{ E}_{\rm FSR\eta}=E_{0\eta}\sum_{plm}\mathbf{ H}^{(p)}_{FR,lm} \mathcal{A}^{(p)}_{\eta lm},
\end{equation}
where
\begin{equation}\label{6.29}
\mathbf{H}^{(p)}_{{\rm FR},lm}=-\frac{{\rm i}}{4\pi k'}\frac{{\rm e}^{{\rm i}
k'r''}}{r}\sum_{\eta'}\mathbf{\hat{u}}_{{\rm S}\eta'}W^{(p)\ast}_{{\rm
S}\eta'lm}(-)^{\eta'+p+l+m}F_{\eta'}(\pi-\vartheta_{\rm S})\;.
\end{equation}
for a sphere on or near the surface, this $\mathbf{ H}$-vector multipole fields with the origin at
$O''$ can be considered as the the mirror image of the source of the original $\mathbf{ H}$ fields.
From the superposition of scattered and reflected-scattered fields, all referred to a common
origin, eqs.\ref{6.28}-\ref{6.30} and \ref{6.33}-\ref{6.29}, we get the field
\begin{align}\label{6.47}
\mathbf{ E}_{\rm Obs\,\eta}=E_{0\eta}\sum_{plm}\mathbf{ H}^{(p)}_{{\rm {Obs}},lm}(\vec{ r},k')
\mathcal{A}^{(p)}_{\eta lm}\;,
\end{align}
with
\begin{equation}\label{6.48}  %(6.37)
\mathbf{H}^{(p)}_{{\rm {Obs}},lm}=-\frac{{\rm i}}{4\pi k'}\frac{{\rm e}^{{\rm i}
k'r'}}{r}\sum_{\eta'}\mathbf{\hat{u}}_{{\rm S}\eta'}W^{(p)\ast}_{{\rm
S}\eta'lm}[1+(-)^{\eta+p'+l+m}F_{\eta'}(\pi-\vartheta_{\rm S}]
\end{equation}
Eqs. (\ref{6.47}-\ref{6.48}) lead us to the definition and derivation of the transition matrix for
a scatterer in the presence of a plane interface\cite{JOSA95,JOSA99,AO99}. The advantages yielded
by the use of the transition matrix is evident if we had to deal with the problem of a random
dispersion of non spherical particles deposited on a plane surface. Moreover, the amplitudes of the
observed field are the key quantities for calculating the radiation force of which we will discuss
later.

\subsection*{S2.1 Optical force in front of a substrate}
In this section we briefly recall our approach to determine the radiation force exerted by a plane
waves, with a definite polarization, on a scatterer (of any shape and composition) placed in a
homogeneous medium of (real) refractive index $n_{\rm m}$. We refer to the geometry sketched in
Fig. \ref{Fig_geom} in which $\Sigma$ is the customary laboratory frame and $\Sigma'$ is a frame of
reference whose axes are parallel to the axes of $\Sigma$  and whose origin $O'$ lies within the
particle. The vector position of $O'$ with respect to $\Sigma$ is $\mathbf{R}_O'$. The conservation
laws applied to the electromagnetic scattering problem leads to the optical force acting on the
particle\cite{Jackson_new, Borghese2007book, Jones2015}:
\begin{equation}\label{1}
\mathbf{F}_{\rm Rad}=r^{\prime2}\int_{\Omega'}\mathbf{\hat{r}}^{\prime}\cdot\langle \mathrm{T}_{\rm
M} \rangle\,{\rm d}\Omega'\;,
\end{equation}

where the integration is over the full solid angle, $r'$ is the radius of a sphere  with center at
$\vec{R}_{O'}$ surrounding the particle, and $\langle \mathrm{T}_{\mathrm M} \rangle$, the averaged
Maxwell stress tensor (MST), describes the mechanical interaction of light with matter. The general
expression of the MST in a medium in the Minkowski form\cite{Jackson_new, Borghese2007book,
Jones2015} is
\begin{equation}
\mathrm{T}_{\rm M} = \mathbf{E}'\otimes\mathbf{D}' + \mathbf{H}'\otimes \mathbf{B}' -
\frac{1}{2}\left(\mathbf{E}'\cdot\mathbf{D}' + \mathbf{H}'\cdot\mathbf{B}' \right) \mathrm{I} \; ,
\end{equation}
where $\mathbf{E}'$ is the electric field, $\mathbf{D}'$ is the electric displacement,
$\mathbf{H}'$ is the magnetic field, $\mathbf{B}'$ is the magnetic induction, evaluated in the
frame $\Sigma'$ as indicated by the prime, $\otimes$ represents the dyadic product, and
$\mathrm{I}$ is the dyadic unit. We assume that all the fields are harmonic, propagating in a
homogeneous, linear, and non-dispersive medium, and depend on time through the factor ${\rm e}^{-i
\omega t}$ that is omitted.
So, we can simplify the expression for the MST by using the complex amplitudes of the fields,
$\textbf{E}' = \textbf{E}'(\textbf{r})$ and $\textbf{B}'=\textbf{B}'(\textbf{r})$,
as\cite{Mishchenko2001, Saija2005,Jones2015}:
\begin{equation}\label{eq:Maxwell_stress_tensor}
\langle  \mathrm{T}_{\mathrm M}\rangle = \frac{\varepsilon_\mathrm{m}}{2}{\rm Re} \left[
\textbf{E}'\otimes\textbf{E}'^{\ast} +
\frac{c^2}{n_\mathrm{m}^2}\textbf{B}'\otimes\textbf{B}'^{\ast} - \frac{1}{2}\left( |\textbf{E}'|^2
+ \frac{c^2}{n_\mathrm{m}^2}|\textbf{B}'|^2 \right) \mathrm{I} \right] ,
\end{equation}
where the fields are the superposition of the incident and of the scattered field. In presence of a
plane surface that separates two homogeneous media with different refractive indexes, the role of
the incident field is played by the exciting field $\mathbf{E}_{\rm E}=\mathbf{E}_{\rm
I}+\mathbf{E}_{\rm R}$ while the superposition of $\mathbf{ E}_{\rm S}$ and $\mathbf{ E}_{\rm SR}$
acts like the observed field due to the presence of particle. It is possible to
simplify\cite{Borghese2007book} equation \eqref{1} since the dyadic products in the expression of
$\langle\mathrm{T}_{\rm M}\rangle$ give a vanishing contribution to the radiative force
\cite{Mishchenko2001, Saija2005}. For these reason, the component of the radiation force along the
direction characterized by the unit vector $\mathbf{\hat{v}}_{\zeta}$ turns out to be
\begin{equation}\label{3}
F_{\rm Rad\,\zeta}=-\frac{1}{4} \varepsilon_{\rm m} r^{\prime2} {\rm Re}
\int_{\Omega'}(\mathbf{\hat{r}}'\cdot\mathbf{\hat{v}}_{\zeta})\bigl[(|\mathbf{E}'_{\rm
Osb}|^2+2\mathbf{E}^{\prime\ast}_{\rm E}\cdot\mathbf{E}'_{\rm Obs})+\frac{c^2}{n_{\rm
m}^2}(|\mathbf{B}'_{\rm Obs}|^2+2\mathbf{B}^{\prime\ast}_{\rm E}\cdot\mathbf{B}'_{\rm
Obs})\bigl]\,{\rm d}\Omega'\;,
\end{equation}
where $\mathbf{E}'_{\rm Obs}$ and $\mathbf{B}'_{\rm Obs}$ are the superposition of the fields
scattered by the particle and the reflected-scattered fields. Obviously, since the exciting field
is a plane wave, the integral \eqref{3} gets no contribution from the terms $\mathbf{E}'_{\rm
E}\cdot\mathbf{E}^{\prime\ast}_{\rm E}$, and $\mathbf{B}'_{\rm E}\cdot\mathbf{B}^{\prime\ast}_{\rm
E}$ that, accordingly, have been omitted. At this stage, using the orthogonality properties of
vector spherical harmonics through which we develop the fields, see eqs. \eqref{6.5}-\eqref{6.6}
and eqs.\eqref{6.28}-\eqref{6.29}, we obtain the Borghese equations for the optical force
components\cite{Borghese2007}:
\begin{align}\label{4}
F_{\rm Rad\,\zeta}=-F^{\rm(Sca)}_{\rm Rad\,\zeta}+F^{\rm(Ext)}_{\rm Rad\,\zeta}\
\end{align}
where
\begin{subequations}\label{6ab}
\begin{align}
F^{\rm(Sca)}_{\rm Rad\,\zeta}&=\frac{\varepsilon_{\rm m}|E_0|^2}{2 k'^2}{\rm Re}\sum_{plm}\sum_{p'l'm'}\mathcal{A}^{(p)\ast}_{lm}\mathcal{A}^{(p')}_{l'm'}i^{l-l'}I^{(pp')}_{\zeta\,lml'm'}\;,\label{6a}\\
F^{\rm(Ext)}_{\rm Rad\,\zeta}&=-\frac{\varepsilon_{\rm m}|E_0|^2}{2 k'^2}{\rm
Re}\sum_{plm}\sum_{p'l'm'}W^{(p)\ast}_{{\rm
E}\,lm}\mathcal{A}^{(p')}_{l'm'}i^{l-l'}I^{(pp')}_{\zeta\,lml'm'}\;,\label{6b}
\end{align}
\end{subequations}
where the matrix elements
\begin{align}
I^{(pp')}_{\zeta\,lml'm'}=\frac{4\pi}{3}\sum_{\mu}Y^{\ast}_{1\mu}(\mathbf{\hat{v}}_{\zeta})\frac{{\rm
i}^{l'-l}}{16\pi^2}\sum_{\eta'}\int Y_{1\mu}(\mathbf{\hat{k}}_{\rm S})W^{(p)}_{{\rm S}{\eta'}
lm}W^{(p')\ast}_{{\rm S}{\eta'} l'm'}\,d\Omega_{\rm S}\;,
\end{align}
can be analytically valuated. We notice that  $F^{\rm(Sca)}_{\rm Rad\,\zeta}$ depends on the
amplitudes $A^{(p)}_{lm}$ of the scattered field only, whereas $F^{\rm(Ext)}_{\rm Rad\,\zeta}$
depends jointly on the amplitudes of the scattered field $A^{(p)}_{lm}$ and on those of the
incident field $W^{(p)}_{{\rm I}\,lm}$. This dependence is analogous to that of the scattering
cross section and of the extinction cross section, respectively, so that $F^{\rm(Sca)}_{\rm
Rad\,\zeta}$ can be somewhat related to scattering properties of the particle, whereas
$F^{\rm(Ext)}_{\rm Rad\,\zeta}$ can be related to its extinction. Similar considerations hold true
also for the radiation torque \cite{Borghese2006, marston1984radiation}.

\begin{figure}
\includegraphics[width=\textwidth]{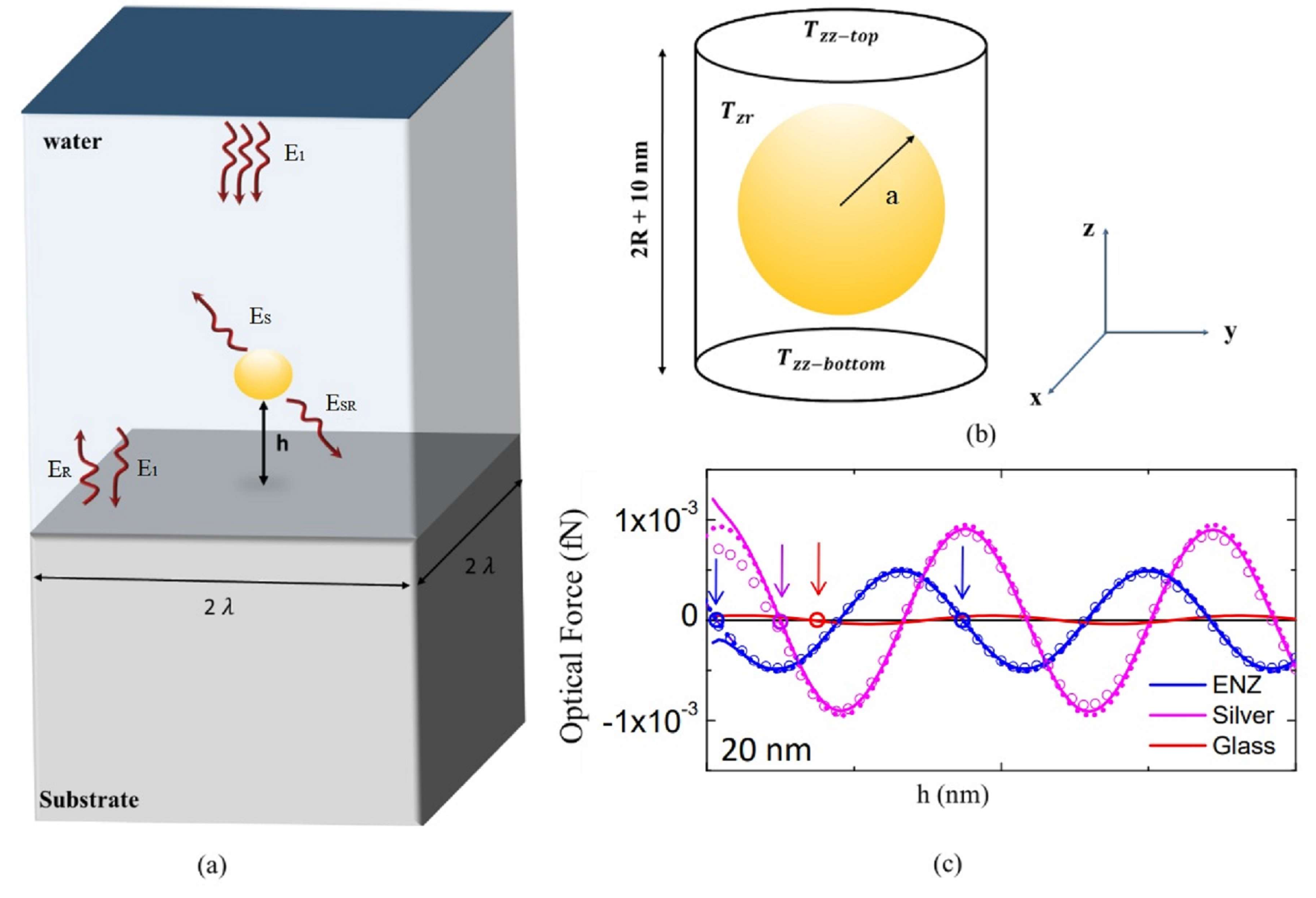}
\caption{Numerical computation of the optical force: (a) Simulation region and the geometry of the
problem. A polystyrene particle is placed above a substrate, the distance $h$ is from the bottom of
the particle to the substrate. The particle is surrounded with water and the incoming plane-wave is
illuminated from the top and propagates in $z$ direction. (b) Calculation of the optical force by
integration of Maxwell's stress tensor on a surface of cylindrical volume surrounding the particle.
The Maxwell's stress tensor consists of information about the total electric and magnetic field
which includes incident field $(E_{\rm I})$, reflected field $(E_{\rm R})$, field scattered by the
particle due to the excitation filed $(E_{\rm S})$, and field scattered by the particle and
reflected by the surface $(E_{\rm RS})$. As we are interested on the optical force in z direction
it is enough to integrate $\mathrm{T}_{\mathrm{M},zz}$ on top and bottom surface and
$\mathrm{T}_{\mathrm{M},rz}$ on the circumferential surface.  (c) [Fig. 2a of the main text]
Numerical computation (circles) versus analytical T-matrix calculation (solid line) and dipole
approximation calculation (dots) of optical force for different distances $h$ above different
substrates (Silver, ENZ, glass).}\label{COMSOL}
\end{figure}

\section*{S3 Finite elements methods}
To evaluate the accuracy of the analytical results, we have computed the optical force on the macro
particle using numerical simulations (Fig. \ref{COMSOL}).  We used software package COMSOL
Multiphysics 5.4 which uses finite element method (FEM) to solve Maxwell's equation and calculate
the optical force. To increase the accuracy of simulation we used periodic boundary condition;
however, the micro-particle radius, $a$, is significantly smaller than the unit cell size, L, to
prevent the mutual coupling between adjacent cells ($a$/L  =0.02) . In order to reduce the
computation time the simulation is run for silicon particle with radius $a$ = 20 nm while the
wavelength of incoming plane wave is $\lambda$=560 nm. The surrounding medium is water. We used
different substrates as reflecting surfaces and compared the computed force for all substrates with
the analytical results. We used silver, glass, layered structure (silver and aluminum dioxide) and
ENZ surfaces. The optical properties used for all surfaces are measured values at $\lambda$=560 nm.
The thickness of all substrates are considerably larger than $\lambda$ to mimic the semi-infinite
medium. The simulation region should be meshed finely specially in three regions: i) the plasmonic
layers (Ag) in the layered structure ii) in the near-field of the substrate $( 0 - \lambda/10 )$ to
capture the near-field effects on the calculated force and iii) the region surrounding the particle
that the force is calculated.

To calculate the force, we used Maxwell's stress tensor. It is known that the total time-averaged
force acting on any material objects can be found by calculating the integral of Maxwell's stress
tensor on any surface that defines a volume containing the objects

\begin{equation}
\mathbf{F}_{\rm Rad}=\langle F(t)\rangle = \int_S \langle \mathrm{T}_{\rm M} (r,t)\rangle \cdot
\mathbf{\hat{n}}\ dS
\end{equation}

where  $\mathrm{T}_{\rm M}$  is the Maxwell's stress tensor calculated based on the total electric
and magnetic fields, $S$ is the surface surrounding the volume containing the object and
$\mathbf{\hat{n}}$  is the unit vector perpendicular to the surface $S$. In the simulation, we
chose a cylinder as a surrounding volume (Fig. \ref{COMSOL}b). In our simulation, we are interested
to calculate the force in $z$ direction consequently $\langle \mathrm{T}_{\mathrm{M},zz}
(r,t)\rangle$ on top and bottom of the cylinder and  $\langle T_{zr} (r,t)\rangle $ on its
circumference should be calculated (Fig. \ref{COMSOL}b). As the normal vector on the top and bottom
of the cylinder has opposite direction, the total time-averaged force is:
\begin{equation}
    \langle F(t)  \rangle =  \int_{S_{Top}} \langle \mathrm{T}_{\mathrm{M},zz} (r,t) \rangle dS -\int_{S_{Bottom}} \langle \mathrm{T}_{\mathrm{M},zz} (r,t)\rangle dS +\int_{S_{Circum}} \langle \mathrm{T}_{\mathrm{M},rz}(r,t)\rangle dS
    \label{F2}
\end{equation}
\\
As integrals in Eq. S\ref{F2} are calculated over the surface of the cylinder, the cylinder is
meshed densely $(\lambda/200)$ to avoid the numerical error. Since the field scattered by the
particle causes no singularity, the height and radius of the cylinder can be as close as possible
to the diameter and radius of the sphere respectively, however the smaller is the cylinder the
denser should be the mesh to capture the intensity of the electric and magnetic field.  To avoid
computational error within reasonable computation time, the height of the cylinder is chosen to be
10 nm bigger than the sphere's diameter and the radius of the cylinder to be 5 nm bigger than the
radius of the sphere. The numerical computation of total force is done for various spacing (h) of
the sphere from different reflecting surfaces (Fig. \ref{COMSOL}a). $h$ is the distance from the
bottom of the sphere to the surface as shown in Fig. \ref{COMSOL}a. The numerical calculation is
done for several scenarios; various spacing $h$, various radii $a$, different substrates, and
particles with different polarizabilities and geometries. The comparison between analytical
solution and numerical analysis for various spacing above various substrates is shown in Fig. 2a.
There is a very good agreement between analytical calculation and numerical simulation.

\section*{S4 Layered metamaterial calculations}

\paragraph{Transfer matrix numerical details.}
The numerical results in Fig. 3 of the main text for reflectance ($R$) versus reflected phase
($\phi$) from the surface of a thin film stack, as described in the caption, are calculated using
the standard transfer matrix approach~\cite{chilwell1984thin}.  The refractive indices at 560 nm
for each material are as follows:  Ag: $0.146 + 3.27i$~\cite{rakic1998optical}; Al$_2$O$_3$:
$1.68$~\cite{boidin2016pulsed}; Au: $0.384 + 2.55i$~\cite{rakic1998optical}; Ge: $3.02 +
2.90i$~\cite{amotchkina2020characterization}; TiO$_2$: 2.43~\cite{siefke2016materials}; glass
substrate: 1.52; water superstrate: 1.33~\cite{daimon2007measurement}.

\paragraph{Experimental comparison.}
The experimental $(\phi,R)$ points shown as green stars in Fig.~\ref{Fig3} are based on six
different fabricated trilayer thin film stack systems (5 trilayers of Al$_2$O$_3$ / Ag / Ge from
top to bottom).  The Ag layer thicknesses were in the range 10-25 nm, with a thin Ge layer (1-3 nm)
underneath to ensure surface wetting.  The Al$_2$O$_3$ thicknesses were systematically varied
between roughly 20 nm and 80 nm across the different systems.  These stacks were deposited on a
glass substrate (Corning Inc.) using electron-beam evaporation for Ge (0.5 $\text{\AA}  /
\text{s}$) and Al$_2$O$_3$ (0.3 $\text{\AA}  / \text{s}$), and thermal evaporation for Ag (0.5
$\text{\AA} / \text{s}$).  All materials were purchased from Kurt J. Lesker.  The sample's
ellipsometric properties (amplitude $\Psi$ and phase difference $\Delta$) were measured in an air
superstrate using a Variable-angle, high-resolution spectroscopic ellipsometer (J. A. Woollam Co.,
Inc, V-VASE) for incident angles $45^\circ$, $50^\circ$, $55^\circ$ and wavelength range $300 -
1000$ nm.  Individual fits to the ellipsometric data for each system yielded best-fit results for
the thicknesses and optical constants, which were then used to estimate the values of $R$ and
$\phi$ at normal incidence with $\lambda = 560$ nm shown in Fig.~3.

\paragraph{Effective medium theory.}
The effective medium theory (EMT) used to calculate the $(\phi,R)$ curves in Fig.~1c and the curve
labeled EMT in Fig.~3 takes the following form.  We consider the interface between a water
superstrate with refractive index $n_0$ and an underlying metamaterial which is a mixture of
dielectric with index $n_d$ and a metal with complex index $\tilde{n}_{\rm M} = n_{\rm M} + i
k_{\rm M}$.  If $f$ is the filling fraction of the metal versus the dielectric, the approximate EMT
permittivity of the metamaterial is given by $\epsilon_\text{EMT} = (1-f) n_d^2 + f \tilde{n}_{\rm
M}^2$.  This allows us to calculate the effective refractive index $n_\text{EMT}$ and extinction
coefficient $k_\text{EMT}$ as:
\begin{equation}\label{emt1}
n_\text{EMT} = \sqrt{\frac{|\epsilon_\text{EMT}|+\epsilon_\text{EMT}}{2}}, \quad k_\text{EMT} =
\sqrt{\frac{|\epsilon_\text{EMT}|-\epsilon_\text{EMT}}{2}}.
\end{equation}
The corresponding complex Fresnel reflection coefficient is given by:
\begin{equation}\label{emt2}
r_\text{EMT} = \frac{n_0 - n_\text{EMT} - i k_\text{EMT}}{n_0 + n_\text{EMT} + i k_\text{EMT}}.
\end{equation}
The associated reflectance $R = |r_\text{EMT}|^2$ and phase $\phi = -\arg{r_\text{EMT}}$.

\bibliographystyle{ieeetr}

\bibliography{main_levitation}

\end{document}